\documentclass[final,3p,times]{elsarticle}

\usepackage{geometry}
\usepackage{graphicx}
\usepackage{amssymb}
\usepackage{booktabs}

\usepackage{caption}
\usepackage{subcaption}
\usepackage{lineno}
\usepackage{array}
\usepackage{float}
\usepackage{tcolorbox}
\usepackage{amsmath,amsthm}
\usepackage{mathrsfs}
\usepackage[colorlinks=true, allcolors=blue]{hyperref}

\usepackage[nodots]{numcompress}

\usepackage{arydshln}

\usepackage{algorithm2e}

\biboptions{sort&compress}
\theoremstyle{remark}

\journal{arXiv}

\begin{document}
\begin{frontmatter}

\title{ViT-Transformer: Self-attention mechanism based constitutive modeling for nonlinear heterogeneous materials}

\author[]{Yijing Zhou\fnref{fn1}}
\author[]{Shabnam J. Semnani\corref{cor2}}
\ead{ssemnani@ucsd.edu}
\date{}
\address{Department of Structural Engineering, University of California, San Diego, La Jolla, CA, 92093, USA}
\cortext[cor2]{Corresponding author}
\fntext[fn1]{Former graduate student}

\begin{abstract}
Multi-scale simulations of nonlinear heterogeneous materials and composites are challenging due to the prohibitive computational costs of high-fidelity simulations. Recently, machine learning (ML) based approaches have emerged as promising alternatives to traditional multiscale methods. 
However, existing ML surrogate constitutive models struggle in capturing long-range dependencies and generalization across microstructures. 
The recent advancements in attention-based Transformer architectures open the door to a more powerful class of surrogate models. Attention mechanism has demonstrated remarkable capabilities in natural language processing and computer vision. In this work, we introduce a surrogate (meta) model, namely ViT-Transformer, using a Vision Transformer (ViT) encoder and a Transformer-based decoder which are both driven by the self-attention mechanism. The ViT encoder extracts microstructural features from material images, while the decoder is a masked Transformer encoder that combines the latent geometrical features with the macroscopic strain input sequence to predict the corresponding stress response.
To enhance training, we propose a random extract training algorithm that improves robustness to sequences of variable length. We design and construct a compact yet diverse dataset via data augmentation, and validate the surrogate model using various composite material images and loading scenarios. Several numerical examples are provided to show the effectiveness and accuracy of the ViT-Transformer model and the training algorithm.

\end{abstract}


\end{frontmatter}

\section{Introduction}

Composites materials are ubiquitous in many engineering applications such as mechanical and aerospace engineering. These materials have a complex microstructure which determines their mechanical behavior, making accurate simulation of composites a challenging task. Moreover, prediction of composite behavior from the microstructural information is necessary for design of composites with enhanced properties. Conventional finite element method (FEM) requires exceedingly fine mesh to capture the material microstructure, which renders high-fidelity computational simulations infeasible. Therefore, various multiscale methods \cite{kanoute2009multiscale,fish2021mesoscopic,semnani2020inelastic} have been developed to allow for capturing the microstructural effects on the macroscopic material response. However, computational multiscale methods, e.g. $FE^{2}$ method \cite{feyel2000fe2}, are prohibitively expensive for path-dependent behavior of composite materials.
Therefore, there is a need for novel approaches which can link the microstructures of nonlinear composite and heterogeneous materials to their macroscopic responses. 

The emergence of artificial intelligence has introduced new opportunities in computational mechanics. Machine learning (ML) based techniques have demonstrated promising results in acceleration of multiscale simulations \cite{logarzo2021smart,xu2020data,kim2024data,zhou2025machine}. One of the common approaches for this purpose is to develop trained ML surrogate models or metamodels that eliminate the need for expensive high-fidelity simulations by learning inherent nonlinear relationships. 
Artificial neural networks (ANNs) represent the earliest and most widely adopted machine learning architecture for surrogate modeling of representative volume elements (RVEs). They have been applied to constitutive responses such as plasticity \cite{jang2021machine}, viscoplasticity \cite{furukawa1998implicit,benabou2021implementation,li2019machine}, and to capture the behavior of polycrystalline metals \cite{ali2019application} and hyperelastic crystal structures \cite{im2021neural}. 

In recent years, attention has turned to more advanced ML architectures, such as graph neural networks (GNNs), which have shown promise both in accelerating finite-element analyses \cite{gulakala2023graph,jiang2023graph,maurizi2022predicting} and in predicting material properties \cite{dai2021graph}.
Recurrent Neural Networks (RNNs) have proven highly effective in capturing path-dependent and rate-dependent material responses \cite{gorji2020potential,he2023machine,li2025molecular}. The hidden state embedded in RNNs serves as an internal memory that preserves information from preceding steps. This feature enables RNNs to excel at handling time-series data and step-wise predictions. To address vanishing-gradient issues, RNN variants such as Gated Recurrent Units (GRUs) and Long Short-Term Memory (LSTM) networks were developed, which have been applied to simulating elasto-plastic \cite{he2023machine,im2021surrogate,guan2023neural,heidenreich2024recurrent,tancogne2021recurrent,dettmer2024framework,heidenreich2024transfer} and viscoelastic \cite{chen2021recurrent} behaviors. Various researchers have applied RNNs to learn path-dependent constitutive laws within homogenization-based multiscale schemes \cite{WU2020113234,ghavamian2019accelerating,logarzo2021smart,haghighi2022single,wu2024self,kim2023surrogate,vijayaraghavan2023data,friemann2023micromechanics,zhou2025machine}. More recently, advanced techniques have been introduced to reduce training data requirements and enhance performance, including approaches based on physics-informed ML \cite{tandale2022physics,maia2023physically,borkowski2022recurrent,deng2024data,qu2021towards,masi2022multiscale,koric2024deep}, neural differential equations \cite{he2024incremental}, and deep operator networks \cite{koric2024deep}.
	
Despite extensive progress, ML-aided simulation techniques face various challenges. RNN-based surrogate models exhibit performance degradation when processing very long sequences because they rely on hidden-state propagation to capture historical information and long-range dependencies. The success of large language models \cite{naveed2025comprehensive} in natural language processing has highlighted the Transformer architecture’s strong capacity for modeling long-range dependencies \cite{NIPS2017_3f5ee243}. Transformer architectures have been employed in various applications such as large-scale physical-systems simulation \cite{zhdanov2025erwin}, crystalline‐material property prediction \cite{NEURIPS2022_6145c70a}, protein structure prediction \cite{jumper2021highly}, gene-expression prediction \cite{avsec2021effective}, and global weather forecasting \cite{lam2023learning}.
Considering their promise in modeling long sequences and  parallelizability, researchers have begun to explore the applications of Transformer-based models in computational mechanics for constitutive modeling \cite{zhongbo2024pre,pitz2024neural}, field prediction \cite{buehler2022fieldperceiver,buehler2022end}, and de novo design of architected materials \cite{yang2021words}. In stress sequence prediction tasks, Zhongbo and Hien \cite{zhongbo2024pre} employed an encoder-only architecture to directly process homogenized strain sequences, while Pitz and Pochiraju \cite{pitz2024neural} integrated Transformers with convolutional neural networks (CNNs) to handle homogenized data derived from RVEs of different geometries. Transformer-based models have been utilized to predict multiscale physical fields and nonlinear material properties \cite{buehler2022fieldperceiver} without reliance on convolutional layers. Additionally, attention mechanisms have been introduced to predict stress fields and fracture patterns \cite{buehler2022end}. 

With the rapid development of computer vision techniques, convolutional neural networks (CNNs) have been applied to the analysis of composites and heterogeneous media to learn the relationship between microstructural features and macroscopic stress responses. For example, Bhaduri et al. \cite{BHADURI2022109879} and Gupta et al. \cite{gupta2023accelerated} mapped fiber layouts directly to stress fields using CNNs to expedite multiscale composite simulations. Subsequent studies extended this framework to elasto-plastic matrices \cite{saha2024prediction} and to viscoplastic polycrystalline materials \cite{khorrami2023artificial}. Additionally, CNNs have been employed to predict effective properties of composite materials \cite{su2023three,chang2022predicting,peng2022ph}.

Existing homogenization-based surrogate models often have limited capability to generalize to unseen microstructures and require re-generating training datasets and re-training models when material microstructure or properties change. Some researchers have proposed using CNNs as feature encoders to extract the microstructural information of composites materials \cite{pitz2024neural}. Recently, Vision Transformers (ViTs) \cite{dosovitskiy2020image} have gained increasing attention in computer vision, and evidence suggests that their performance can surpass CNN-based models with less computational requirements for training \cite{dosovitskiy2020image,liu2021swin,touvron2021training}. The adoption of ViT architectures within computational mechanics, however, remains largely unexplored.
	
In this work, we introduce ViT-Transformer, a surrogate model that combines a Vision Transformer encoder responsible for capturing microstructural information from RVE images with a Transformer decoder that predicts macroscopic stresses using the encoder’s latent features and an input sequence of macroscopic strains. The resulting model shows high accuracy and strong generalization across diverse composite microstructures. To further improve robustness, we devise a novel training algorithm that enhances the model’s ability to handle sequences of varying lengths without modifying the underlying dataset. We also provide a detailed data generation protocol and present a set of experiments to validate the performance of the proposed model in unseen scenarios. The proposed model demonstrate a great performance even though the number of trainable parameters and the size of the training dataset are smaller compared to those reported in previous works for composite homogenization.
The primary contributions of this work are summarized below:
\begin{itemize}
	\item This work proposes a novel surrogate constitutive model, namely, the ViT-Transformer, for composite and heterogeneous materials. This approach incorporates the self-attention mechanism from two aspects: A ViT encoder to extract the microstructural features of the composite, and a Transformer decoder for stress prediction. This surrogate model generalizes well across composites with differing microstructures.
	\item We propose a new random extract training algorithm to enhance the generalization capability of the Transformer surrogate model when processing strain sequences of varying lengths. This algorithm avoids the need for modifying the training dataset to sequences of equal length for batch training, and is more efficient than the existing padding approach in the literature.
    \item We demonstrate that the surrogate model based on self-attention mechanism outperforms the conventional GRU-based architecture when processing long strain sequences.
	\end{itemize}

	The structure of this paper is as follows. Section \ref{sec:problem_statement} reviews the microscale RVE problem and homogenization. Section \ref{sec:methodology} provides an overview of the self-attention mechanism, the original Transformer model, and the ViT-Transformer architecture proposed in this work. In Section \ref{sec:training_strategies}, we discuss the training data generation and augmentation approach, a novel random extract training algorithm, the learning rate adjustment strategy, and the details of the training process. Section \ref{sec:results} compares the performance of our Transformer-based decoder with GRU, and provides a wide range of numerical examples to validate the present methodology across different microstructures and loading scenarios.

\section{Problem statement}\label{sec:problem_statement}

\subsection{Microscale problem}\label{sec:RVE}

The first step in multiscale modeling of heterogeneous material is to set up the boundary value problem at the microscale. We consider the 3D  microscopic structure of a composite material with long continuous fibers. Representative volume elements (RVEs) contain two phases, namely, fiber and matrix. In this work we use a carbon fiber reinforced metal matrix composite for demonstration which has widespread engineering applications such as aerospace and the automotive industry. The reinforcement is an intermediate-modulus carbon fiber modeled as linearly elastic material with a Young’s modulus of 324 GPa, and a Poisson’s ratio of 0.1 and a tensile strength of 7 GPa. The matrix is specified as an aluminum alloy modeled with J2 plasticity. The Young’s modulus, Poisson’s ratio, and hardening modulus of the matrix are assumed as 71.7 GPa, 0.33, and 0.013. 

\begin{figure}[h!]
    \centering
    \includegraphics[width=10cm]{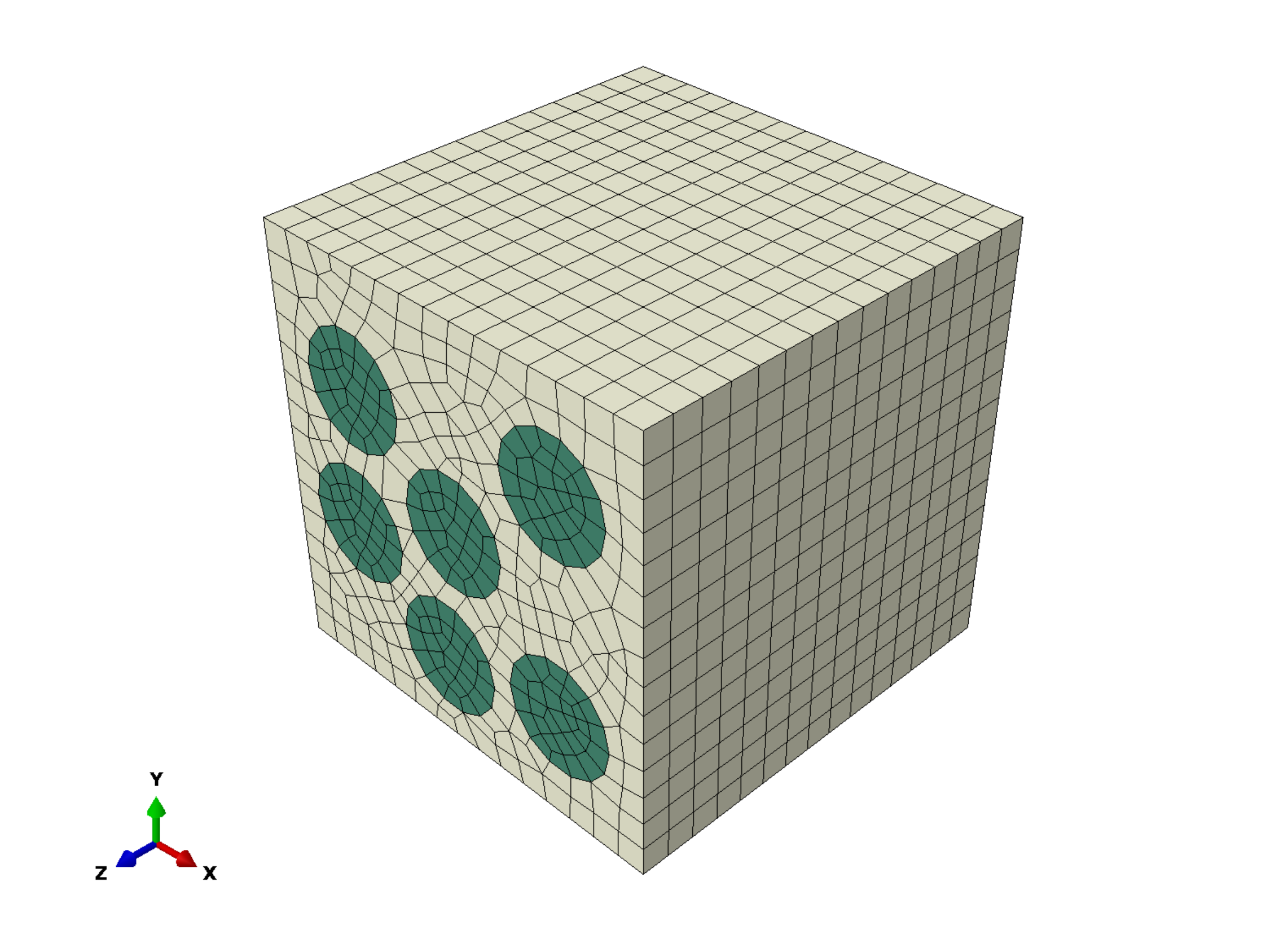}
    \caption{An example RVE model generated by the preprocessing script with randomly distributed fibers. Fibers are located along the Z direction.}
    \label{fig:rve_model}
\end{figure}

To streamline the analysis, we have developed a preprocessing script which takes the number of RVEs, number of fibers, radius of fibers, and size of the RVE models as inputs to automatically generate a large number of RVE models.  Details of the preprocessing algorithm and script are included in \cite{zhou2025machine}. After creating the geometry, the algorithm assigns material properties, generates the mesh, prescribes a sequence of macroscopic strains, and enforces periodic boundary conditions (PBC) on each RVE. An example RVE generated with this script is shown in Figure \ref{fig:rve_model}.

To enforce periodicity on the opposite boundaries, we apply PBC to the RVEs. Displacement $\mathbf{u}$ is written as
\begin{equation}
\mathbf{u}(\mathbf{X}, \mathbf{x}) = \boldsymbol{E}(\mathbf{X})\cdot \mathbf{x} + \mathbf{u'}(\mathbf{x}) \, ,
\label{eq:bc1}
\end{equation} 
in which $\mathbf{u'}$ represents periodic displacement fluctuations. $\mathbf{x}$ and $\mathbf{X}$ are the microscopic and macroscopic position vectors, respectively. Using Equation \ref{eq:bc1}, the strain field is obtained as

\begin{equation}
\boldsymbol{\varepsilon} = \boldsymbol{E} + \Tilde{\boldsymbol{\varepsilon}} \, ,
\label{eq:bc2}
\end{equation}
where $\mathbf{E}$ and $\tilde{\boldsymbol{\varepsilon}}$ denote the macroscopic strain and microscopic fluctuations, respectively. Volume average of the microscopic fluctuations is zero, that is,

\begin{equation}
\frac{1}{V}\int_{V}^{}\tilde{\boldsymbol{\varepsilon}} \, \mathrm{d}V = 0 \, ,
\label{eq:bc3}
\end{equation}
in which V is the total RVE volume.
The microscale boundary value problem is written as 

\begin{equation}
\left\{\begin{array}{c}
\nabla_{x}.\boldsymbol{\sigma}(\mathbf{x}) = 0 \, ,\\
\boldsymbol{\sigma} = \mathbf{F}(\mathbf{x}, \boldsymbol{\varepsilon}) \, ,\\
\mathbf{u}_i - \mathbf{u}_j = \boldsymbol{E} \cdot (\mathbf{x}_i - \mathbf{x}_j) \, ,
\end{array}\right. 
\qquad
\label{eq:bc5}
\end{equation}
where $\mathbf{x}_i$, $\mathbf{x}_j$, $\mathbf{u}_i$, and $\mathbf{u}_j$, denote the positions and displacements of each pair of points on the opposite sides of the RVE surface, respectively. $\mathbf{F}$ represents a function that defines the micro-scale constitutive relationships between stress and strain, and $\boldsymbol{\sigma}$ is the microscopic stress field. 

To impose periodic boundary conditions in ABAQUS, we set nine reference points with unit displacements for every RVE model to apply the macroscopic strain series to the nodes on the surfaces of the RVEs. The following equation constraint is used to constrain corresponding nodes and reference points: 

\begin{equation}
C_{1}\cdot DOF_{1} + C_{2}\cdot DOF_{2} + C_{3}\cdot DOF_{3} = 0 \, ,
\label{eq:rm1}
\end{equation}
in which $DOF_{1}$, $DOF_{2}$, and $DOF_{3}$ are the corresponding degrees of freedom (displacements) of the first node, the opposite node, and the reference point, respectively. $C_1$, $C_2$ and $C_3$ are appropriate coefficients to impose linear constraints. Here we have $C_{1}=1$, $C_{2}=-1$, while $C_{3}$ depends on the macroscopic strain following Equation \ref{eq:bc5}.

\subsection{Homogenization}\label{sec:homogenization}

One of the most important aspects of multiscale simulations is transfer of information between the microscopic and macroscopic scales. We use the periodic homogenization technique \cite{tikarrouchine2018three,semnani2020inelastic,choo2021anisotropic} to define the equivalent macroscopic stress as

\begin{equation}
\boldsymbol{\Sigma} = \frac{1}{V}\int_{V}^{}\boldsymbol{\sigma} \, \mathrm{d}V \, .
\label{eq:ho1}
\end{equation}
Equations \ref{eq:bc2} and \ref{eq:bc3} directly yield the macroscopic strain $\boldsymbol{E}$ as
\begin{equation}
\boldsymbol{E} = \frac{1}{V}\int_{V}^{}\boldsymbol{\varepsilon} \, \mathrm{d}V \, .
\label{eq:ho2}
\end{equation}
We discretize the integrals in Equations \ref{eq:ho1} and \ref{eq:ho2} to perform numerical implementation as follows:
\begin{equation}
\boldsymbol{\Sigma} = \frac{1}{V}\sum_{i=0}^{N_{q}}v_i\boldsymbol{\sigma}_i \, ,
\label{eq:ho3}
\end{equation}

\begin{equation}
\boldsymbol{E} = \frac{1}{V}\sum_{i=0}^{N_{q}}v_i\boldsymbol{\varepsilon}_i \, ,
\label{eq:ho4}
\end{equation}
where $\boldsymbol{\sigma}_i$ and $\boldsymbol{\varepsilon}_i$ denote the microscopic stress and strain tensors at integration point $i$, respectively. $N_{q}$ is the number of integration or quadrature points in the RVE. $v_i$ is the volume corresponding to integration point $i$.  
We have developed a postprocessing script in ABAQUS CAE (see \cite{zhou2025machine} for details) which extracts the simulation outputs from ODB files and applies Equations \ref{eq:ho3} and \ref{eq:ho4} to calculate the macroscopic stresses and strains.

\section{Methodology}\label{sec:methodology}

\subsection{Self-attention mechanism}\label{sec:SA}

Self-attention mechanism is the core component of the Transformer architecture, which is used to learn the dependency relationships in long sequences \cite{NIPS2017_3f5ee243}. The self-attention mechanism can directly learn the dependency relationship between any two positions in the input sequence, regardless of the distance between them.  Having shown outstanding performance in the fields of NLP and CV, another advantage of the Transformer architecture is its amenability to parallelization. The details of self-attention mechanism are described below.

Let matrix $\mathbf{Z} \in \mathbb{R}^{n \times d}$ represent the embedded input sequence, where $n$ is the sequence length and $d$ is the embedding (or input) dimension. Each input vector $\mathbf{z}_i$ represents the $i$-th position in the sequence. Matrix $\mathbf{Z}$ is given to the attention mechanism as input and undergoes three different linear transformations to obtain the query ($\mathbf{Q}$), key ($\mathbf{K}$), and value ($\mathbf{V}$) matrices:

\begin{equation}
\mathbf{Q} = \mathbf{Z} \mathbf{W}^Q, \quad \mathbf{K} = \mathbf{Z} \mathbf{W}^K, \quad \mathbf{V} = \mathbf{Z} \mathbf{W}^V  \, ,
\label{eq:Self_Attention_1}
\end{equation}
in which $\mathbf{W^Q}, \mathbf{W^K}, \mathbf{W^V} \in \mathbb{R}^{d \times d_k}$ are trainable parameter matrices. $d_k$  is typically set to $ d/h$, where h is the number of heads in the multi-head attention mechanism (see Section \ref{sec:MSA}). Matrix form of Equation \ref{eq:Self_Attention_1} reads

\begin{equation}
\mathbf{Q} = \begin{bmatrix}
\mathbf{z}_1 \\
\mathbf{z}_2 \\
\vdots \\
\mathbf{z}_n
\end{bmatrix}
\begin{bmatrix}
w_{11}^Q & w_{12}^Q & \cdots & w_{1d_k}^Q \\
w_{21}^Q & w_{22}^Q & \cdots & w_{2d_k}^Q \\
\vdots & \vdots & \ddots & \vdots \\
w_{d1}^Q & w_{d2}^Q & \cdots & w_{dd_k}^Q
\end{bmatrix}
= \begin{bmatrix}
\mathbf{q}_1 \\
\mathbf{q}_2 \\
\vdots \\
\mathbf{q}_n
\end{bmatrix} _{n \times d_k}\, ,
\label{eq:Self_Attention_Query}
\end{equation}

\begin{equation}
\mathbf{K} = \begin{bmatrix}
\mathbf{z}_1 \\
\mathbf{z}_2 \\
\vdots \\
\mathbf{z}_n
\end{bmatrix}
\begin{bmatrix}
w_{11}^K & w_{12}^K & \cdots & w_{1d_k}^K \\
w_{21}^K & w_{22}^K & \cdots & w_{2d_k}^K \\
\vdots & \vdots & \ddots & \vdots \\
w_{d1}^K & w_{d2}^K & \cdots & w_{dd_k}^K
\end{bmatrix}
= \begin{bmatrix}
\mathbf{k}_1 \\
\mathbf{k}_2 \\
\vdots \\
\mathbf{k}_n
\end{bmatrix}_{n \times d_k} \, ,
\label{eq:Self_Attention_Key}
\end{equation}

\begin{equation}
\mathbf{V} = \begin{bmatrix}
\mathbf{z}_1 \\
\mathbf{z}_2 \\
\vdots \\
\mathbf{z}_n
\end{bmatrix}
\begin{bmatrix}
w_{11}^V & w_{12}^V & \cdots & w_{1d_k}^V \\
w_{21}^V & w_{22}^V & \cdots & w_{2d_k}^V \\
\vdots & \vdots & \ddots & \vdots \\
w_{d1}^V & w_{d2}^V & \cdots & w_{dd_k}^V
\end{bmatrix}
= \begin{bmatrix}
\mathbf{v}_1 \\
\mathbf{v}_2 \\
\vdots \\
\mathbf{v}_n
\end{bmatrix}_{n \times d_k} \, .
\label{eq:Self_Attention_Value}
\end{equation}

After obtaining the three matrices $\mathbf{Q}$, $\mathbf{K}$, and $\mathbf{V}$, we compute the attention score. For this purpose, first the dot product of the query and key matrices are calculated as 

\begin{equation}
	\mathbf{Q} \mathbf{K}^T = \begin{bmatrix}
		\mathbf{q}_1 \\
		\mathbf{q}_2 \\
		\vdots \\
		\mathbf{q}_n
	\end{bmatrix}
	\begin{bmatrix}
		\mathbf{k}_1^T & \mathbf{k}_2^T & \cdots & \mathbf{k}_n^T
	\end{bmatrix}
	= \begin{bmatrix}
		\mathbf{q}_1 \cdot \mathbf{k}_1 & \mathbf{q}_1 \cdot \mathbf{k}_2 & \cdots & \mathbf{q}_1 \cdot \mathbf{k}_n \\
		\mathbf{q}_2 \cdot \mathbf{k}_1 & \mathbf{q}_2 \cdot \mathbf{k}_2 & \cdots & \mathbf{q}_2 \cdot \mathbf{k}_n \\
		\vdots & \vdots & \ddots & \vdots \\
		\mathbf{q}_n \cdot \mathbf{k}_1 & \mathbf{q}_n \cdot \mathbf{k}_2 & \cdots & \mathbf{q}_n \cdot \mathbf{k}_n
	\end{bmatrix} _{n \times n}\, ,
	\label{eq:Q_KT}
\end{equation}
where each element $\mathbf{q}_i \cdot \mathbf{k}_j$ represents the similarity between the $i$-th query vector and the $j$-th key vector. To avoid the problem of large dot product values leading to gradient vanishing or explosion, we divide the dot product result by a scaling factor $\sqrt{d_k}$. Scaling helps keep the dot product results within a reasonable range, making the subsequent softmax operation more stable. Subsequently, we apply the softmax normalization function row-wise to obtain the attention score (weight) matrix $\boldsymbol{\alpha}$ containing attention weights for each query vector over all key vectors. 

\begin{equation}
\boldsymbol{\alpha} = \text{softmax}\left(\frac{\mathbf{Q} \mathbf{K}^T}{\sqrt{d_k}}\right) \, ,
\label{eq:attention-score}
\end{equation}
where the softmax function of a vector $\boldsymbol{z}$  is defined in Equation \ref{eq:softmax}  and converts the scaled similarity values into a probability distribution, ensuring that the scores for each query vector sum to 1.
\begin{equation}
\text{softmax}(\boldsymbol{z})_i = \frac{e^{z_i}}{\sum_{j=1}^n e^{z_j}}
\label{eq:softmax}
\end{equation}
The attention score  $\alpha_{ij}$ captures how much each key vector $j$ is relevant for query vector $i$, thus determining the degree to which each query vector should focus on each key vector when computing the final output.
Finally, we use the attention score matrix to perform a weighted sum of the value matrix $\mathbf{V}$, yielding the final output

\begin{equation}
\text{Attention}(\mathbf{Q}, \mathbf{K}, \mathbf{V}) = \boldsymbol{\alpha} \mathbf{V} =  \begin{bmatrix}
\alpha_{11} & \alpha_{12} & \cdots & \alpha_{1n} \\
\alpha_{21} & \alpha_{22} & \cdots & \alpha_{2n} \\
\vdots & \vdots & \ddots & \vdots \\
\alpha_{n1} & \alpha_{n2} & \cdots & \alpha_{nn}
\end{bmatrix}
\begin{bmatrix}
\mathbf{v}_1 \\
\mathbf{v}_2 \\
\vdots \\
\mathbf{v}_n
\end{bmatrix}
= \begin{bmatrix}
\sum_{j=1}^n \alpha_{1j} \mathbf{v}_j \\
\sum_{j=1}^n \alpha_{2j} \mathbf{v}_j \\
\vdots \\
\sum_{j=1}^n \alpha_{nj} \mathbf{v}_j
\end{bmatrix}_{n \times d_k}
\label{eq:Attention_matrix}
\end{equation}

\subsection{Multi-head self-attention mechanism}\label{sec:MSA}

Multi-head self-attention mechanism is commonly used in Transformer-based architectures, which is an extension of self-attention mechanism by performing $h$ self-attention processes in parallel and concatenating the output. While the single-head attention mechanism only uses one attention function,  the multi-head attention mechanism uses multiple attention functions (i.e. multiple heads), each with an independent query, key, and value matrix. Through this approach, different heads can focus on  different parts of the input sequence, thereby improving the model's ability in capturing richer features and relationships.

The self attention mechanism is extendend to multiple heads, each with its own set of linear transformations, as

\begin{equation}
\text{MultiHead}(\mathbf{Q}, \mathbf{K}, \mathbf{V}) = \begin{bmatrix}
\text{head}_1 ,
\text{head}_2 ,
\hdots ,
\text{head}_h
\end{bmatrix}
\mathbf{W}^O
\label{eq:Multi_head_attention_matrix}
\end{equation}

where

\begin{equation}
\text{head}_i = \text{Attention}(\mathbf{Q}_i, \mathbf{K}_i, \mathbf{V}_i) = \text{Attention}(\mathbf{Z} \mathbf{W}_i^Q, \mathbf{Z} \mathbf{W}_i^K, \mathbf{Z} \mathbf{W}_i^V) 
\label{eq:Multi_head_attention_head}
\end{equation}
In Equation \ref{eq:Multi_head_attention_head}, $\textbf{W}_{i}^Q, \textbf{W}_{i}^K , \textbf{W}_{i}^V \in \mathbb{R}^{d \times d_k}$ represent the transformation matrices of the $i$-th head to convert the input sequence into the query vector $\textbf{Q}_i$,  the key vector $\textbf{K}_i$, and the value vector $\textbf{V}_i$, respectively. After concatenating the outputs of all heads, a linear projection is performed using $\textbf{W}^O  \in \mathbb{R}^{h d_k \times d}$ to obtain the final output.

\subsection{Overview of the original Transformer model }\label{sec:original}

\begin{figure}[h!]
    \centering
    \includegraphics[width=0.9\textwidth]{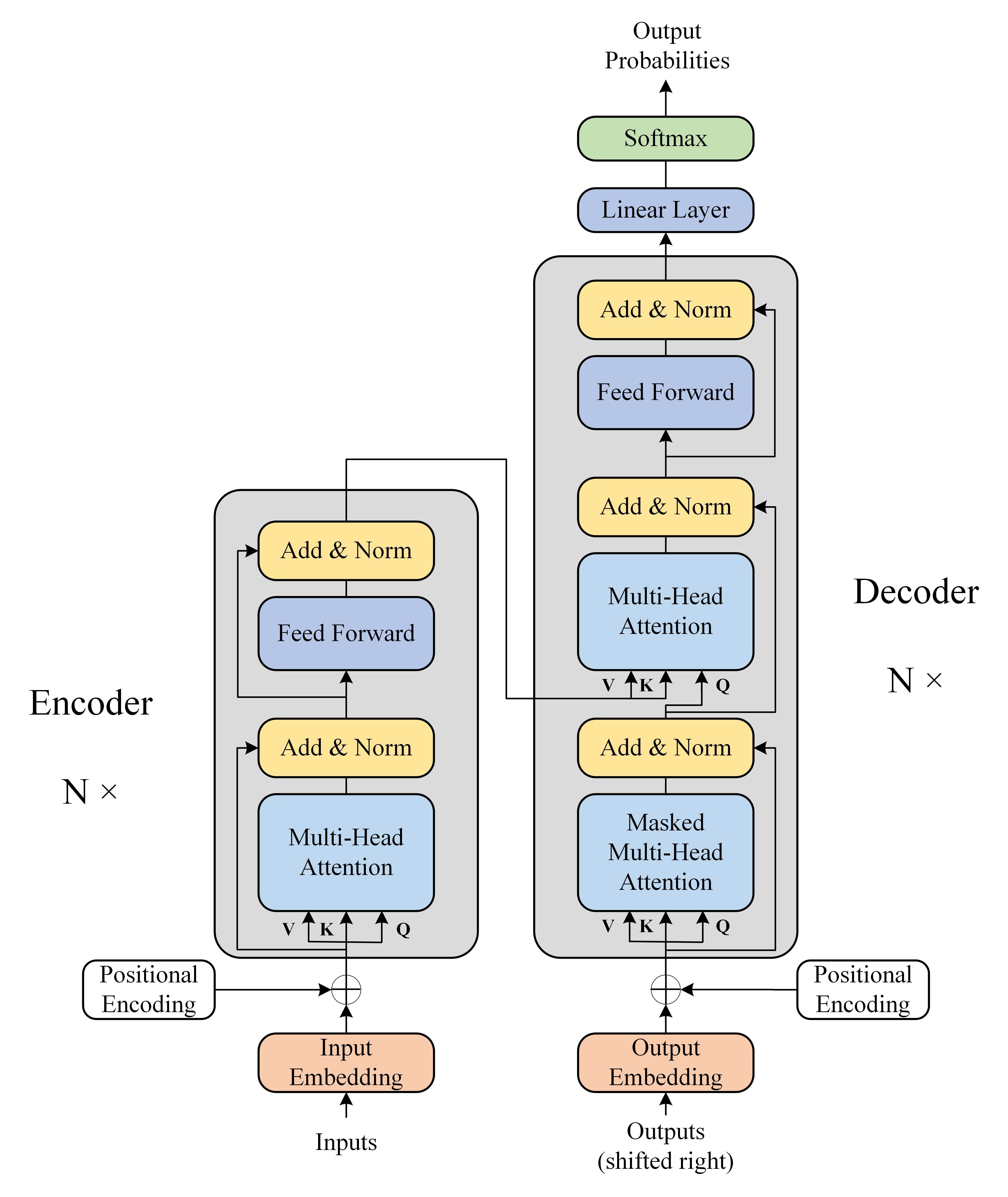}
    \caption{Architecture of the original Transformer model in \cite{NIPS2017_3f5ee243}.}
    \label{fig:Transformer_original_model}
\end{figure}

The original Transformer architecture proposed by Vaswani et al. \cite{NIPS2017_3f5ee243} is composed of 6 encoder layers and 6 decoder layers stacked together, as shown in Figure \ref{fig:Transformer_original_model}. Each layer of the encoder contains a multi-head self-attention mechanism and a two-layer feedforward neural network, both of which use layer normalization \cite{ba2016layer}. The activation function is typically ReLU or GELU, and here we select ReLU.
In each layer of the decoder, the model sequentially connects a masked multi-head self-attention mechanism, a multi-head self-attention mechanism, and a feedforward neural network. A residual connection is used for each sublayer along with layer normalization; that is, the output of each sublayer is derived as

 \begin{equation}
 	\mathbf{o} = \text{LayerNorm}(\mathbf{x}_{in} + \text{Sublayer}(\mathbf{x}_{in}))
 	\label{eq:Residual_LayerNorm}
 \end{equation}
 where $\text{Sublayer}(\mathbf{x}_{in})$ represents the function implemented by the sublayer itself (multi-head attention layer or feed-forward layer) with input $\mathbf{x}_{in}$, and $\text{LayerNorm}$ is a function for layer normalization. The fully connected feed-forward sublayers in the encoder and decoder consist of two linear transformations with an activation function (here ReLU) in between the two layers as
 \begin{equation}
\text{FFN}(\mathbf{x}_{in}) = \text{ReLU}(\mathbf{x}_{in} \mathbf{W}_1 + \mathbf{b}_1)\mathbf{W}_2 + \mathbf{b}_2 \, ,
\label{eq:FFN_layer}
\end{equation}
 where \(\mathbf{W}_1, \mathbf{b}_1, \mathbf{W}_2,\) and \(\mathbf{b}_2\) denote the weights and biases of the linear layers.
 
 It should be pointed out that the output of the encoder is used as the key and value of the multi-head attention layer in the decoder, while its query is from the masked multi-head attention layer of the decoder itself. In the Transformer architecture, masking is implemented by adding a large negative constant to the attention scores corresponding to invalid or future positions before the softmax operation. This ensures that the masked positions receive near-zero attention weights after normalization, effectively preventing leakage of information from the subsequent positions of the sequence into the preceding positions. Masking is thus critical for enforcing autoregressive behavior in decoding and handling variable-length sequences.

The Transformer model also requires input embedding, output embedding, and positional encoding structures. In NLP, input embedding is the process of converting the vocabulary in an input sequence into high-dimensional vector representations which help capture the semantic information. Output embedding is similar to input embedding, which is the process of converting vocabulary in the target sequence into high-dimensional vector representations, which is used in the decoder. Due to the lack of built-in sequential information in Transformers, they also need to explicitly incorporate positional information into vocabulary embeddings through positional encoding. Positional encoding adds unique positional information to the embedding of vocabulary, enabling the model to recognize the relative and absolute positions of vocabulary in the sequence.

The positional encoding tensor \( \mathbf{P} \in \mathbb{R}^{n \times d} \) for each position \( pos \in \{0, 1, 2, \ldots, n\} \) in the input sequence and each dimension index $i=0,1,...,(d/2-1)$ in the embedding dimension $d$ is defined as 

\begin{equation}
\mathbf{P}(pos, 2i) = \sin\left(\frac{pos}{10000^{\frac{2i}{d}}}\right)
\label{eq:Positional_encoding_sin} \, ,
\end{equation}

\begin{equation}
\mathbf{P}(pos, 2i+1) = \cos\left(\frac{pos}{10000^{\frac{2i}{d}}}\right)
\label{eq:Positional_encoding_cos} \, .
\end{equation}

The embedded representation \( \mathbf{Z} \) is obtained as
\begin{equation}
\mathbf{Z} = \mathbf{X}\mathbf{W}_e + \mathbf{D}_e + \mathbf{P} \, ,
\label{eq:Positional_encoding}
\end{equation}
in which \( \mathbf{X} \in \mathbb{R}^{n \times d_{in}} \) is the input of the model. \( \mathbf{W}_e \in \mathbb{R}^{d_{in} \times d} \) and \( \mathbf{D}_e \in \mathbb{R}^{n \times d} \) denote the embedding weight and bias matrices of the model, respectively. \( d_{in} \) indicates the input dimension. For example, when the model input is strain sequences in 3D, \( d_{in}=6 \) representing the number of strain components. For image inputs, \( d_{in}=p^2 c \), where $p$, and $c$ denote the image patch size and number of channels.

\subsection{ViT-Transformer}\label{sec:vit}

\begin{figure}[h!]
    \centering
    \includegraphics[width=\textwidth]{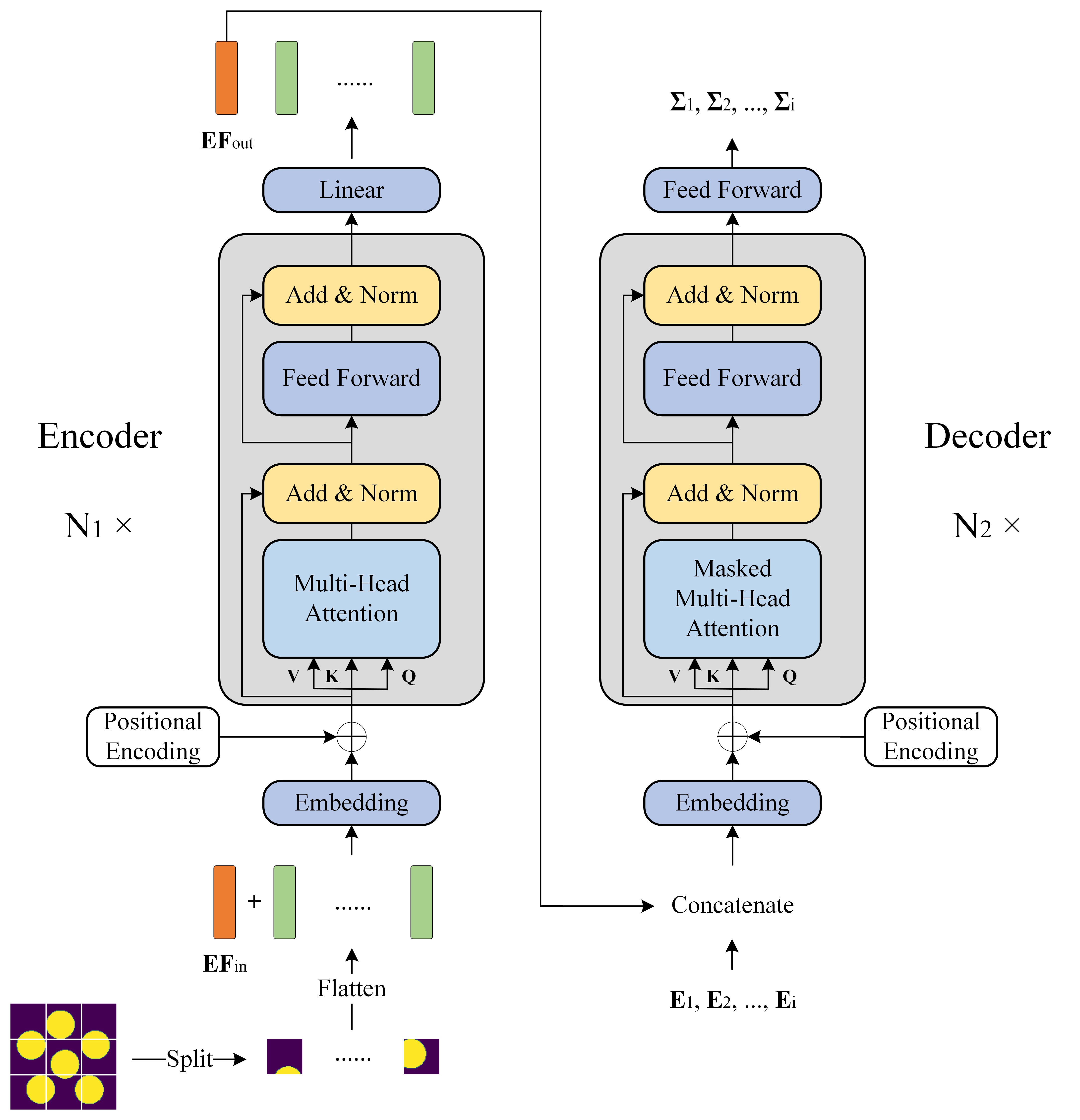}
    \caption{Architecture of the ViT-Transformer.}
    \label{fig:ViT-Transformer}
\end{figure}

In this section, we introduce ViT-Transformer, an architecture consisting of an encoder and a decoder both of which are based on the self-attention mechanism, as shown in Figure \ref{fig:ViT-Transformer}. The encoder is a Vision Transformer that extracts microstructural features of the composite material. Specifically, microstructure images of the composite RVE, taken on planes perpendicular to the fiber direction, are fed to the encoder. Each image is split into $N$ patches with size $p$ which are arranged sequentially in a row-wise manner from left to right. Splitting an image into non-overlapping patches enables the ViT to capture long-range spatial dependencies. The patches are subsequently flattened into vectors and mapped into an embedding space. 

A feature extractor vector $\textbf{EF}_{in} \in \mathbb{R}^{d}$ is concatenated to the left-hand side of the sequence of embedded patches. The embedded input sequence is obtained as
\begin{equation}
\mathbf{Z} = \left[ \textbf{EF}_{in}; \mathbf{x}_1 \mathbf{W}_e; \mathbf{x}_2 \mathbf{W}_e; \cdots; \mathbf{x}_N \mathbf{W}_e \right] + \mathbf{D}_e + \mathbf{P}
\label{eq:Positional_encoding_ViT_encoder}
\end{equation}
in which $\mathbf{x}_i$ is the flattened representation of the $i$-th patch and \( \mathbf{P} \in \mathbb{R}^{(N + 1) \times d} \) is the positional encoding tensor.
The feature extractor's input values could in general be arbitrary or even trainable. In this work, we set all elements of $\textbf{EF}_{in}$ to $\frac{1}{2}$. Since the input images encode fiber regions as 1 and matrix regions as 0, this choice streamlines the feature representation and expedites convergence. 
Positional encodings are applied before the sequence enters the attention layer to distinguish the relative positions of the patches. The architecture of the encoder is similar to the original Transformer model described in Section \ref{sec:original}, comprising an unmasked multi-head self-attention layer and a position-wise feed-forward network, each using residual connections and layer normalization. The final step of the encoder is a linear projection layer without an activation function to compress the dimensionality of the encoder’s output.

The decoder of the ViT-Transformer combines the latent features extracted by the encoder with the input macroscopic strain sequence to predict stresses. For this purpose, the first position of the encoder output, the feature extractor with microscopic information, $\textbf{EF}_{out} \in \mathbb{R}^{d_{out}}$, is isolated, broadcast, and concatenated to every time step of the input strain sequence to be processed: 
\begin{equation}
\mathbf{EF} = \mathbf{1}_n \mathbf{EF}_{out}^\top \, ,
\label{eq:broadcast}
\end{equation}

\begin{equation}
\mathbf{X}_{decoder} = \left[ \mathbf{E}; \;\; \mathbf{EF} \right] \in \mathbb{R}^{n \times (d_s + d_{out})} \, ,
\label{eq:concatenated}
\end{equation}
in which $\mathbf{1}_n \in \mathbb{R}^{n}$ is an all-ones vector. $\mathbf{EF} \in \mathbb{R}^{n \times d_{out}}$, $\mathbf{E} \in \mathbb{R}^{n \times d_s}$, and $\mathbf{X}_{decoder}$ denote the broadcasted feature extractor, macroscopic strain sequence, and the input of the decoder, respectively. $d_{out}$ is the dimension of encoder output features, and $d_s$ is the dimension of strain vector which is equal to 6 for the three-dimensional case in this work.

Before entering the decoder, the concatenated sequence  undergoes embedding and positional encoding by applying a linear transformation and adding the positional encoder $\mathbf{P}$, similar to Equation \ref{eq:Positional_encoding_ViT_encoder}. Unlike the original Transformer model, our decoder does not include a cross-attention module since stress prediction only requires strain data together with the latent features. Accordingly, the decoder retains only a masked multi-head self-attention layer and feed-forward layer. When predicting stress at the current step, the causal mask prevents the model from accessing future strain values. This implementation is necessary since in a typical finite element simulation, load is applied incrementally and strain at future time steps is not known. Finally, the decoder output goes through a feed-forward network with LeakyReLU activation function to predict the macroscopic stress sequence, $\mathbf{\Sigma}$, as 
\begin{equation}
\mathbf{\Sigma} = \text{LeakyReLU}(\mathbf{y} \mathbf{W}_{out1} + \mathbf{b}_{out1})\mathbf{W}_{out2} + \mathbf{b}_{out2} \, ,
\label{eq:decoder_out}
\end{equation}
where $\mathbf{y}$ denotes the decoder output, and \(\mathbf{W}_{out1}, \mathbf{b}_{out1}, \mathbf{W}_{out2},\) and \(\mathbf{b}_{out2}\) represent the weight matrices and bias vectors of the first and second layers, respectively.

Similar to the original Transformer architecture, the encoder and the decoder of our model are both realized as deep stacks of multiple blocks, with the encoder comprising ${N}_{1}$ layers and the decoder comprising ${N}_{2}$ layers. Figure \ref{fig:ViT-Transformer} illustrates the architecture of ViT-Transformer in detail. Hyperparameters of ViT-Transformer in this work are summarized in Table \ref{tab:ViTT_HP}. Hyperparameter tuning is perfromed by empirically selecting an initial configuration and fine-tuning each hyperparameter individually while holding all others constant to determine the final set of hyperparameters.
The total number of trainable parameters for this architecture is 2.26 million.

\begin{table}[h!]
\centering
\caption{Hyperparameters of ViT-Transformer.}
\label{tab:ViTT_HP}
\renewcommand{\arraystretch}{1.1}
\begin{tabular}{l l l l}
\hline 
\textbf{Hyperparameter} & \textbf{Encoder} & \textbf{Decoder} \\[5pt]\hline 
Patch Size & 8 & -- \\[5pt]
Input Size & 64 & 70 \\[5pt]
Embedding Dimension & 80 & 120 \\[5pt]
Number of Heads & 2 & 10 \\[5pt]
Transformer Feedforward Size & 500 & 800 \\[5pt]
Number of Encoder (or Decoder) Layers & 6 & 6 \\[5pt]
Dimension of Output FFN Layers & -- & 720 \\[5pt]
Number of Output FFN Layers & -- & 2 \\[5pt]
Output Size & 64 & 6 \\[5pt]\hline
\end{tabular}
\end{table}

\section{Model training}\label{sec:training_strategies}

\subsection{Dataset generation and augmentation}\label{sec:dataset_generation_and_augmentation}

The dataset in this work comprises macroscopic strains, macroscopic stresses, and microstructure images. Each 3D composite RVE considered in this work (e.g. Figure \ref{fig:rve_model}) is represented by its 2D cross-sectional image in which fiber and matrix regions are assigned values of 1 and 0, respectively. The strain dataset is generated with the random-walk-based sampling algorithm proposed in our previous work \cite{zhou2025machine}, where random strain increments are drawn from a uniform distribution in a pre-defined range for each strain component.
The stress dataset is obtained by homogenizing high-fidelity RVE simulation results (see Section \ref{sec:homogenization}). The image dataset of RVE cross-sections are randomly produced using a geometry generator that places fibers at random locations within the matrix. The full original dataset contains 1000 samples, partitioned into five groups of 200 samples each. Within each group, the number of fibers is identical, whereas their spatial positions are randomly distributed. The length of the strain and stress sequences is set to 100. The resolution of the images is 128 by 128, and the number of channels is 1.

\begin{figure}[h!]
	\centering
	\includegraphics[width=0.9\textwidth]{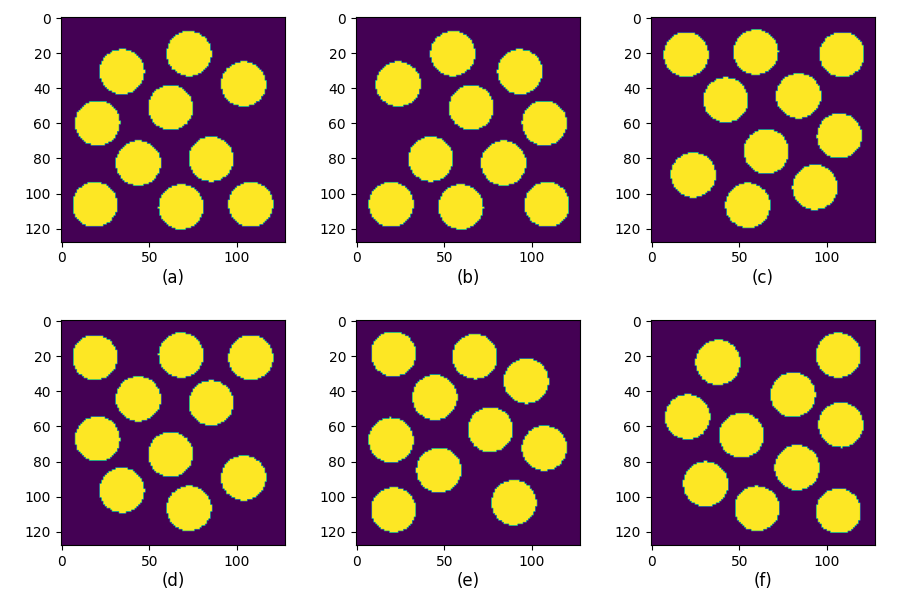}
	\caption{An example of dataset augmentation: a) the original image, b) flip along the y-axis, c) rotate 180 degrees clockwise, d) flip along the x-axis, e) rotate 90 degrees clockwise, and f) rotate 270 degrees clockwise.}
	\label{fig:Dataset_Augmentation}
\end{figure}

We further expand the dataset through data augmentation, as shown in Figure \ref{fig:Dataset_Augmentation}. Each original image is flipped about the $x$-axis and $y$-axis and rotated clockwise by $90^{\circ}$, $180^{\circ}$, and $270^{\circ}$. The transformed images are then incorporated into the dataset. During these operations, the corresponding components of the stress and strain sequences are transformed accordingly. After augmentation, the dataset comprises 6000 samples (1200 samples per group).  Finally, we appended 1000 samples with pure matrix (J2 elasto-plastic) and 1000 samples with pure fiber (linear elastic) material, accompanied, respectively, by 1000 all-zero and 1000 all-one images. Including these two groups of data enables the model to capture the intrinsic characteristics of elastic and elasto-plastic constitutive behaviors. We note that the generated strain sequences vary randomly within and across the groups mentioned above. Details of the dataset are summarized in Table \ref{tab:Dataset}. Out of the 8000 generated samples, 7500 are allocated to the training set and the remaining 500 to the test set.

\begin{table}[h]
	\centering
	\caption{Composition of the dataset.}
	\label{tab:Dataset}
	\renewcommand{\arraystretch}{1.1}
	\begin{tabular}{c c c c }
		\hline
		\textbf{Group No.} & \textbf{No. of Fibers} & \textbf{Radius of Fibers} & \textbf{No. of Samples} \\[3pt]\hline
		1 & 3 & 0.175 & 1,200 \\[3pt]
		2 & 5 & 0.13 & 1,200 \\[3pt]
		3 & 6 & 0.135 & 1,200 \\[3pt]
		4 & 8 & 0.125 & 1,200 \\[3pt]
		5 & 10 & 0.1 & 1,200 \\[3pt]
		6 (pure matrix) & 0 & 0 & 1,000 \\[3pt]
		7 (pure fiber) & 0 & 0 & 1,000 \\[3pt]\hline
		Total &  & & 8,000 \\[3pt]\hline
	\end{tabular}
\end{table}

\subsection{Random extract training algorithm}

In order to enable the Transformer model to process sequences of different lengths, researchers often use datasets containing sequence data with various lengths for model training. To facilite batching in NLP, special padding tokens (commonly [PAD]) are added to the end of shorter sequences to ensure all sequences within a batch have a uniform length. To prevent the model from being misled by the padding tokens during training, masking techniques are employed. A masking matrix marks the padding positions as invalid to exclude them during the attention mechanism and loss calculations. When computing attention weights, the masking matrix assigns a negative infinity weight to the padding positions, effectively nullifying their contribution to the results. Another approach consists of placing sequences of similar lengths within the same batch and then pad or dynamically adjust the padding length based on the longest sequence in the current batch to make the training process more efficient.

Although the above padding approach solves the problem of unequal sequence lengths during training, the process is relatively cumbersome. Contrary to NLP where sequence length plays a crucial role in shaping semantics, in the context of material constitutive modeling, learning the constitutive relationships of materials is independent of the lengths of strain or stress sequences. Therefore, we propose an alternative approach, namely, the random extract training (RET) algorithm. 

First, our data generator creates a dataset of macroscopic stresses and strains. The sequences in these datasets are of equal length $l$, which is larger than the desired sequence length. Before beginning training on each batch, the algorithm generates a random integer index, $n$ between $l$ and a preset minimum length $l_{min}$.
Subsequently, only the first $n$ steps of all sequences in the current batch are extracted for training. During training, we randomly regenerate $n$ for each batch. In addition, when our training reaches the penultimate epoch, we set $n = l$ to improve model performance in processing sequences with the maximum length. This method is detailed in the Algorithm \ref{alg:RE_Training}. 

\SetKwComment{Comment}{/* }{ */}
\RestyleAlgo{ruled}
\begin{algorithm}
\caption{Random Extract Training Algorithm}\label{alg:RE_Training}
\KwIn{ Strain sequence dataset $X_{train}$,  stress dataset $Y_{train}$,  image dataset $X_{train2}$, training parameters, defined model, testing dataset, defined loss function loss\_func}
\KwResult{ViT-Transformer model training is completed}
\BlankLine
Initialize ViT-Transformer model\;
Initialize the necessary tensors\;
$l_{min} \gets$ minimum length of sequence\;
$l \gets$ length of sequences in $X_{train}$\;
$N_{epochs} \gets$ number of epochs\;
$L_{dataset} \gets$ size of the training dataset\;
$b \gets$ batch size\;
\For{epoch = 0 \textbf{to} $N_{epochs}$}{
    \For{i = 0 \textbf{to} ($L_{dataset}$ / b)}{
    $random\_index \gets$ random.randint($l_{min}, l$)\;
    \If{$epoch == N_{epochs} - 2$}{
        $random\_index \gets l$\;
    }
        $X_{train\_i} \gets X_{train}[i*b : (i+1)*b, 0 : random\_index, :]$\;
        $Y_{train\_i} \gets Y_{train}[i*b : (i+1)*b, 0 : random\_index, :]$\;
        $X_{train\_2\_i} \gets X_{train\_2}[i*b : (i+1)*b, :, :, :]$\;
        $prediction \gets$ Model$(X_{train\_i}, X_{train\_2\_i})$\;
        $loss \gets$ loss\_func$(prediction, Y_{train\_i})$\;
        Save training loss\;
        Compute and save testing loss\;
        Perform backpropagation and update model\;
    }
    Adjust the learning rate\;
}
\end{algorithm}

Training the Transformer with Algorithm \ref{alg:RE_Training}  simplifies data generation and training process. We note that this algorithm is effective for learning stress strain relationships since the current step of the stress sequence is only related to past information. However, it is not applicable in natural language processing since extracting an incomplete sentence may lead to semantic loss.

\subsection{Learning rate adjustment}\label{sec:learnrate}

Dynamically adjusting the learning rate is crucial when training Transformer models. This method can accelerate convergence, save computational resources, stabilize training and reducing overfitting. The warmup adjustment method \cite{NIPS2017_3f5ee243} is a commonly used learning rate adjustment strategy for Transformer-based architectures. In this approach, the learning rate is gradually increased in the initial stage of training to help prevent training instability caused by excessive gradients during the early stages of training. In this work, we design a learning rate adjustment strategy by combining the warmup method and learning rate schedulers, as detailed in Algorithm \ref{alg:lr_Adjustment}. Here we set $n_2=20$, $n_3=10$, and the reduction coefficients are selected as $\gamma_2=0.8$ and $\gamma_3= 0.8$.

\SetKwComment{Comment}{/* }{ */}
\RestyleAlgo{ruled}
\begin{algorithm}
\caption{Learning Rate Adjustment Algorithm}\label{alg:lr_Adjustment}
\KwIn{Initial learning rate $lr_0$, minimum learning rates $lr_{min1}$ and $lr_{min2}$, scaling factors $\gamma_2$ and  $\gamma_3$, number of epochs in the warmup stage $n_1$, epoch steps $n_2$ and $n_3$, number of epochs $N_{epochs}$}
\KwResult{Learning rate ($lr$) is adjusted}
\BlankLine
Initialize the Transformer model\;
$\gamma_1 \gets$ 1.0965\;
$n_1 \gets$ 50\;
$lr \gets lr_0$\;
\For{epoch = 0 \textbf{to} $N_{epochs}$}{
    Train for current epoch\;
    \uIf{epoch $< n_1$ }{
        $lr \gets \gamma_1 * lr$\;
    }
    \uElseIf{epoch $\geq n_1$  \textbf{and} lr $> lr_{min1}$}{
        \If{$(epoch - n_1 ) \% n_2$ == 0}{
            $lr \gets \gamma_2 * lr$\;
        }
    }
    \Else{
        \If{The training loss does not decrease after $n_3$ consecutive epochs \textbf{and} lr $> lr_{min2}$ }{
            $lr \gets \gamma_3 * lr$\;
        }
    }
}
\end{algorithm}

In the warmup stage of our learning rate adjustment algorithm, we set a small initial learning rate and scale it by a factor of 100 within $n_1=50$ epochs. After the warmup stage, we multiply the learning rate by the preset reduction factor $\gamma_2$ after every $n_2$ number of epochs as long as the learning rate is still larger than minimum learning rate $lr_{min1}$. Subsequently, whenever the training loss does not decrease for $n_3$ consecutive epochs,  we multiply the learning rate by the preset reduction coefficient $\gamma_3$. Once learning rate becomes smaller than $lr_{min2}$, it will no longer be adjusted.

\subsection{Training process}\label{sec:training}

In this work, we use the Adaptive Moment Estimation (Adam) algorithm to train our model. The dataset needs to be normalized before training to avoid convergence issues and model bias due to different magnitues of strain and stress components.We use the Min-Max scaler as follows:

\begin{equation}
{\Sigma'}_{i,j} = \frac{\Sigma_{i,j} - \Sigma_{\text{min}}}{\Sigma_{\text{max}} - \Sigma_{\text{min}}}, \quad {E'}_{i,j} = \frac{E_{i,j} - E_{\text{min}}}{E_{\text{max}} - E_{\text{min}}} \, ,
\label{eq:minmax}
\end{equation}
where $\Sigma_{\text{min}}$ and $\Sigma_{\text{max}}$ are the minimum and maximum values of stress across all samples, time steps, and components in the dataset. Similarly,
$E_{\text{min}}$ and $E_{\text{max}}$ are the minimum and maximum values of strain.
$\Sigma_{i,j}$ and $E_{i,j}$ represent component $j$ of the $i$th original stress and strain data points, and $\Sigma'_{i,j}$ and $E'_{i,j}$ are their normalized counterparts.  

We select the mean square error (MSE) as the loss function

\begin{equation}
\text{MSE} = \frac{1}{NT} \sum_{n=1}^{N} \sum_{t=1}^{T} \left\| \boldsymbol{\Sigma}_{n,t} - \hat{\boldsymbol{\Sigma}}_{n,t} \right\|_{2}^{2} \, , 
\label{eq:mse}
\end{equation}
in which $\boldsymbol{\Sigma}_{n,t}$ and $\hat{\boldsymbol{\Sigma}}_{n,t}$ are the true (reference) and predicted stress tensors of  the $n$th sample at time step $t$. $N$ and $T$ denote the total number of samples and time steps, respectively.

\begin{figure}[h!]
    \centering
    \includegraphics[width=12cm]{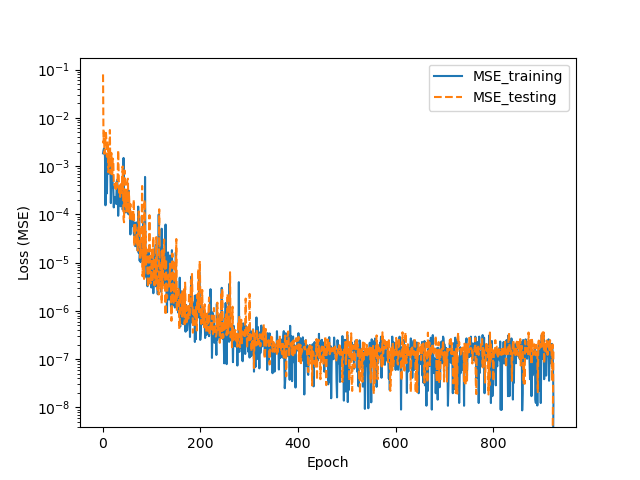}
    \caption{Training and test loss curves versus epochs.}
    \label{fig:Loss_curves}
\end{figure}

The model is trained using PyTorch on a single NVIDIA V100 SXM2 GPU. Using an initial learning rate of $1 \times 10^{-5}$, a batch size of 20, and 900 training epochs, the total training time was approximately 6,600 seconds (about 1.8 hours). Figure \ref{fig:Loss_curves} shows the training and test loss curves versus epochs, showing that the training MSE is about $1 \times 10^{-7}$ at the end of the training process. We observe that both training and test losses have converged, which indicates that the model is trained well. 

\section{Results}\label{sec:results}

In this section, we conduct a series of experiments to validate the proposed ViT-Transformer surrogate modeling approach and demonstrate its performance. We use the following relative error metric to compare the results of our model with reference solutions:

\begin{equation}
E(\boldsymbol{\hat{\Sigma}}) = \frac{ {\parallel\boldsymbol{\hat{\Sigma}} - \boldsymbol{\Sigma}_{ref}\parallel}_2 }{ {\parallel\boldsymbol{\Sigma}_{ref}\parallel}_2 } \,,
\label{eq:relative-error}
\end{equation}
in which $E(\boldsymbol{\hat{\Sigma}})$, $\boldsymbol{\hat{\Sigma}}$ and $\boldsymbol{\Sigma}_{ref}$ denote, respectively, the relative error in stress, the predicted stress vector series (for all time steps), and corresponding reference values for the stress series. In the following numerical examples, the reported relative errors are calculated using Equation \ref{eq:relative-error} for each sequence and averaged over the number of series.

\subsection{Performance of ViT-Transformer}

\subsubsection{Performance on the testing dataset}

\begin{figure}[h]
  \centering
  \begin{subfigure}[c]{0.15\textwidth}
      \centering
      \includegraphics[width=\textwidth]{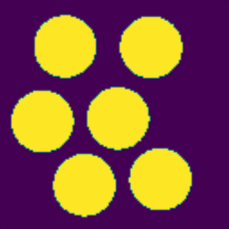}
  \end{subfigure}
  \hfill
  \begin{subfigure}[c]{0.84\textwidth}
      \centering
      \includegraphics[width=\textwidth]{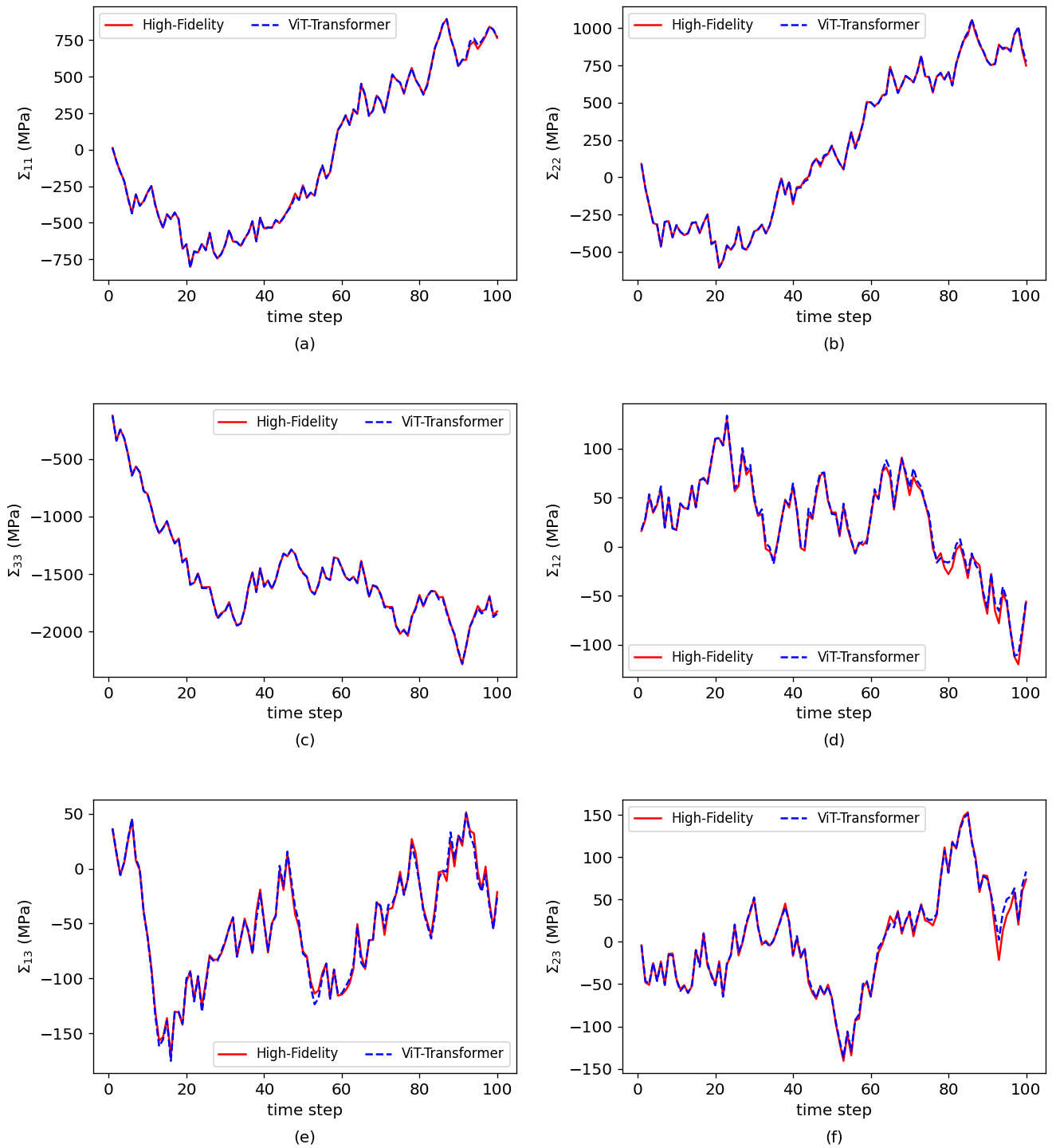}
  \end{subfigure}
  \caption{Comparison of the true and predicted macroscopic stress corresponding to a sample from the test dataset. The 2D cross-sectional image of the microstructure is shown on the left. (a-f) Predicted and reference macroscopic stress components versus time steps. Reference values are shown in red and blue dashed lines show the predicted values.}
  \label{fig:Test_in_testing_dataset}
\end{figure}

We first evaluate the performance of the trained model on the test dataset, generated as discussed in Section \ref{sec:dataset_generation_and_augmentation}. 
Figure \ref{fig:Test_in_testing_dataset} shows an example of the model’s performance on a sample from the test dataset for all six stress components, corresponding to the microstructure shown in Figure \ref{fig:Test_in_testing_dataset}. Figures \ref{fig:Test_in_testing_dataset}(a-f) compare the homogenized stress predicted by ViT-Transformer and the reference solution obtained from homogenization of the high-fidelity simulation. We observe that the predicted values match the references values well across all time steps. The total relative error of the model calculated on the entire test dataset is 1.16 \%, obtained using Equation \ref{eq:relative-error} for each sequence and averaged over the number of sample stress series.

\subsubsection{Performance on unseen loading patterns}\label{sec:unseen_load}

\begin{figure}[h]
    \centering
    \includegraphics[width=12cm]{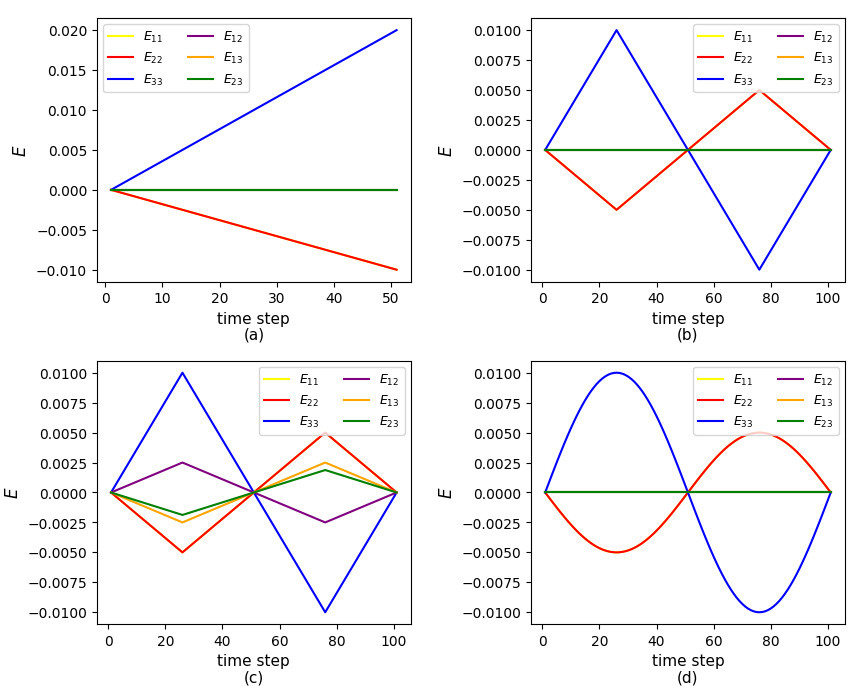}
    \caption{Macroscopic strain protocols used for model validation: (a) monotonic load with zero shear strain, (b) linear cyclic load with zero shear strains, (c) linear cyclic load with non-zero shear strains, and (d) sinusoidal load with zero shear strains.}
    \label{fig:triaxial_states}
\end{figure}

In this section, we evaluate the performance of the ViT-Transformer surrogate model on loading patterns that are not included in the training or testing datasets. The composite microstructures are newly generated at random. 
For this purpose, four new loading protocols are considered for macroscopic strain: monotonic (Figure \ref{fig:triaxial_states}a), linear cyclic with zero shear strains (Figure \ref{fig:triaxial_states}b), linear cyclic with non-zero shear strain (Figure \ref{fig:triaxial_states}c), and sinusoidal cyclic (Figure \ref{fig:triaxial_states}d) protocols. The macroscopic strains from these loading protocols are applied to high-fidelity simulations of the RVEs. Subsequently, the homogenized macroscopic stresses are calculated and used as reference values. The predicted macroscopic stress values are obtained from the ViT-Transformer model using the given macroscopic strain series and RVE microstructure as input.

\begin{figure}[h]
  \centering
  \begin{subfigure}[c]{0.15\textwidth}
      \centering
      \includegraphics[width=\textwidth]{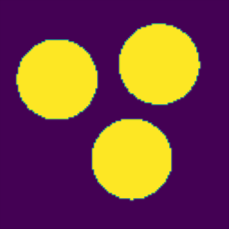}
      \label{fig:Test_in_unseen_1_1}
  \end{subfigure}
  \hfill
  \begin{subfigure}[c]{0.84\textwidth}
      \centering
      \includegraphics[width=\textwidth]{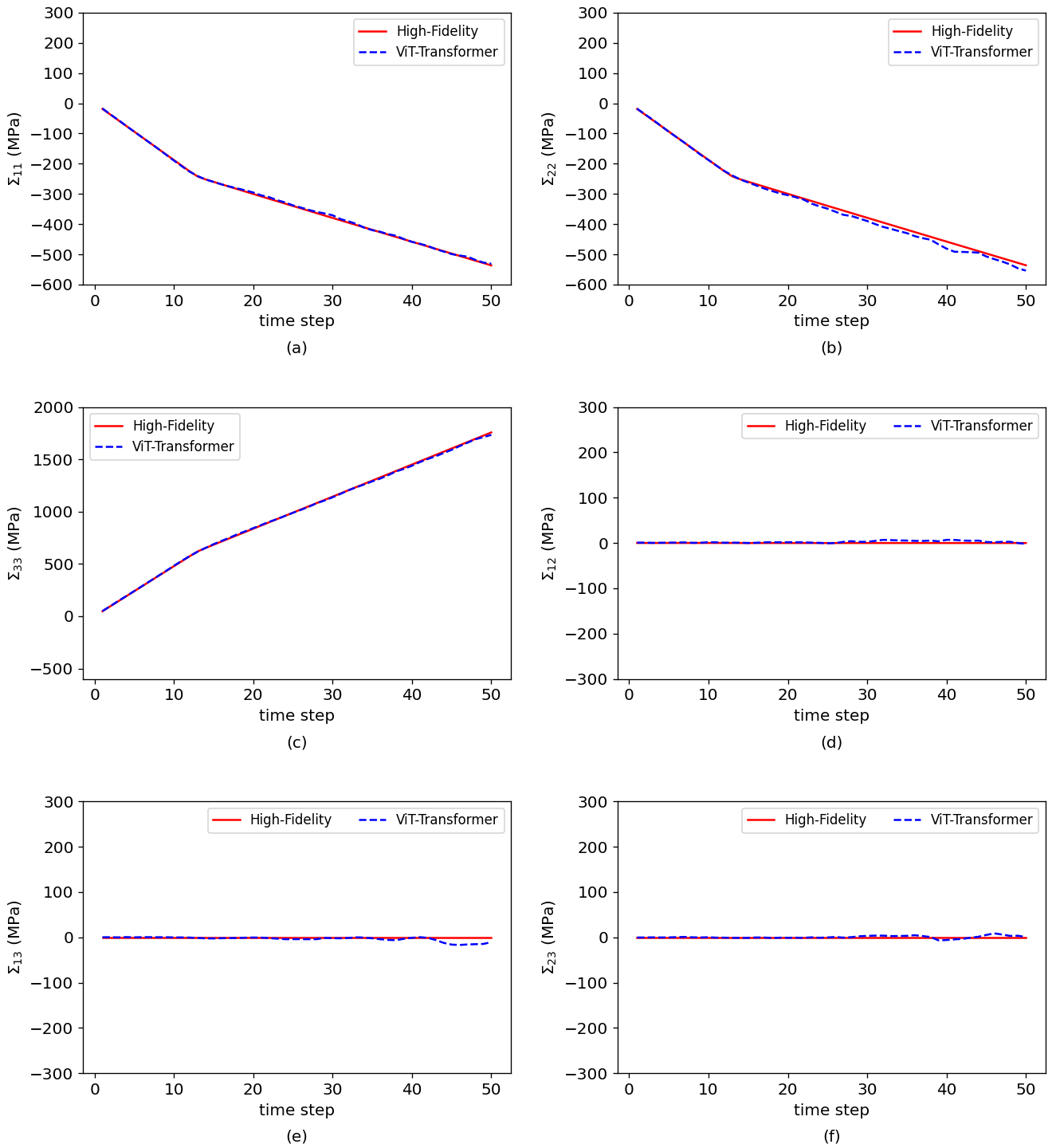}
      \label{fig:Test_in_unseen_1_2}
  \end{subfigure}
  \caption{Comparison of the predicted and true stresses corresponding to the unseen monotonic strain loading protocol shown in Figure \ref{fig:triaxial_states}a. The 2D cross-sectional image of the microstructure is shown on the left. (a-f) Predicted and true macroscopic stresses versus time steps. Reference values are shown in red and blue dashed lines show the predicted values.}
  \label{fig:Test_in_unseen_1}
\end{figure}

The monotonic loading protocol consists of 50 time steps (Figure \ref{fig:triaxial_states}a). $E_{33}$ varies monotanically from 0 to 0.02 with fixed increments of 0.0004, while $E_{11}$ and $E_{22}$ decrease from 0 to -0.01 with increments of -0.0002.  The shear components remain zero. The geometrical features in this example are the same as Group 1 in Table \ref{tab:Dataset}, and the positions of fibers are randomly generated and unseen (Figure \ref{fig:Test_in_unseen_1}). The predicted and reference values of macroscopic stress are illustrated in Figure \ref{fig:Test_in_unseen_1} for all components, and the relative error in this case is 1.25\%. 

\begin{figure}[h]
  \centering
  \begin{subfigure}[c]{0.15\textwidth}
      \centering
      \includegraphics[width=\textwidth]{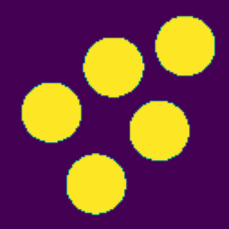}
      \label{fig:Test_in_unseen_2_1}
  \end{subfigure}
  \hfill
  \begin{subfigure}[c]{0.84\textwidth}
      \centering
      \includegraphics[width=\textwidth]{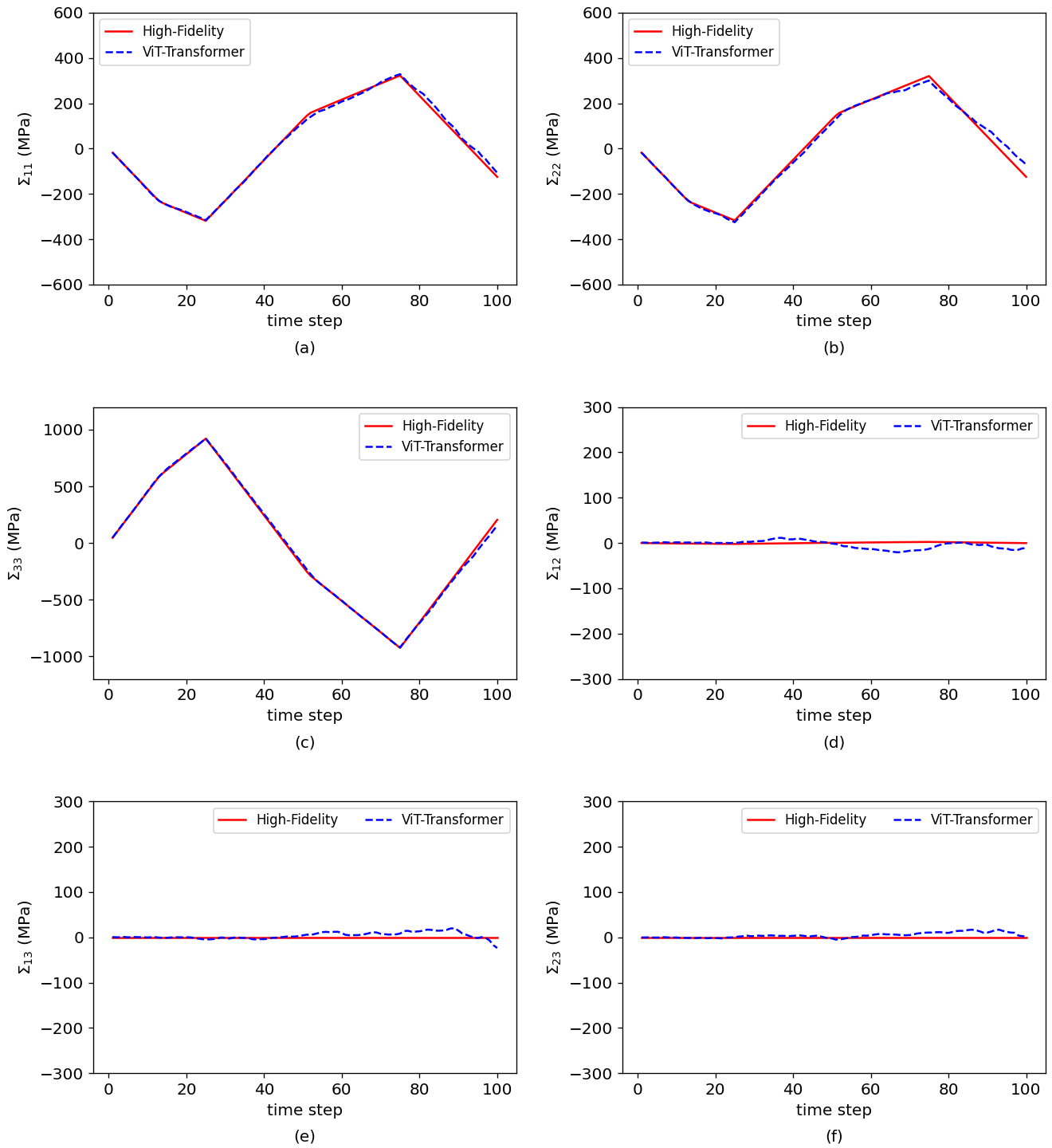}
      \label{fig:Test_in_unseen_2_2}
  \end{subfigure}
  \caption{Comparison of the predicted and true stresses corresponding to the unseen linear cyclic strain loading protocol shown in Figure \ref{fig:triaxial_states}b. The 2D cross-sectional image of the microstructure in this example is shown on the left. (a-f) Predicted and true macroscopic stresses versus time steps. Reference values are shown in red and blue dashed lines show the predicted values.}
  \label{fig:Test_in_unseen_2}
\end{figure}

\begin{figure}[h]
    \centering
    \includegraphics[width=12cm]{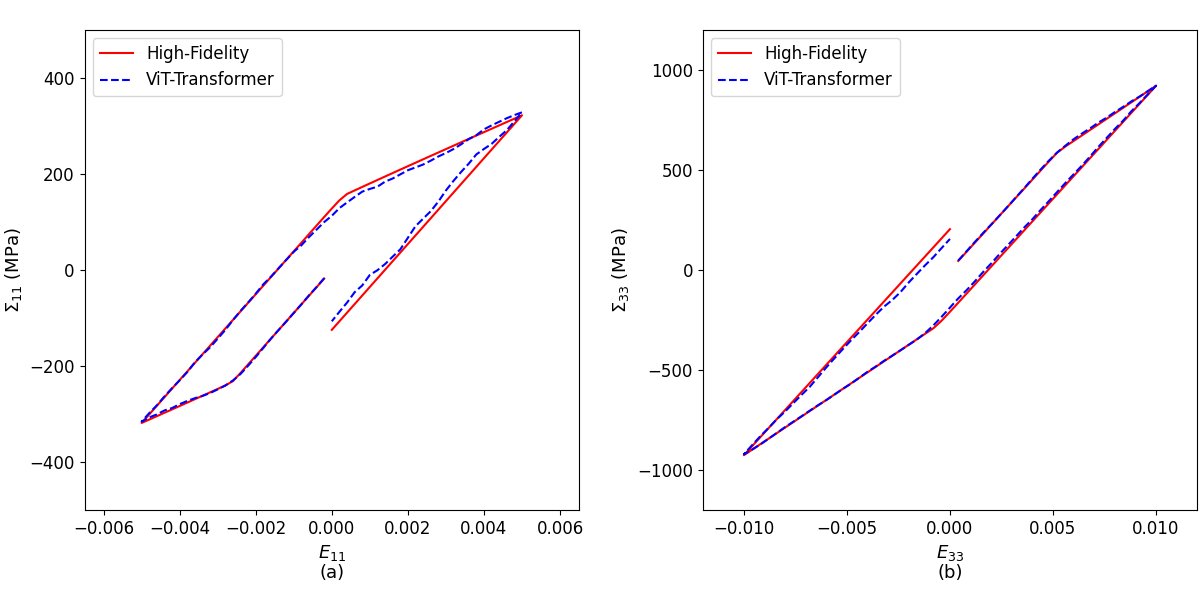}
    \caption{The reference and predicted macroscopic stress versus strain corresponding to the linear cyclic loading with zero shear as shown in Figure \ref{fig:triaxial_states}b.}
    \label{fig:Test_in_unseen_2_3}
\end{figure}

The linear cyclic protocol with zero shear strains consists of 100 time steps (Figure \ref{fig:triaxial_states}b). $E_{33}$ reaches the peak value of $\pm$0.01 with fixed increment size of $\pm$0.0004, while $E_{11}$ and $E_{22}$ reach a peak strain of $\pm$0.005 with $\pm$0.0002 increment size. The geometrical features in this example are the same as Group 2 in Table \ref{tab:Dataset}. The positions of fibers are randomly generated and unseen (Figure \ref{fig:Test_in_unseen_2}). 
Figures \ref{fig:Test_in_unseen_2} and \ref{fig:Test_in_unseen_2_3} show the stress time history and stress versus strain behavior of the ViT-Transformer model, respectively. The relative error in this case is 4.87\%.

\begin{figure}[h]
  \centering
  \begin{subfigure}[c]{0.15\textwidth}
      \centering
      \includegraphics[width=\textwidth]{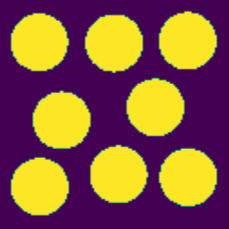}
      \label{fig:Test_in_unseen_3_1}
  \end{subfigure}
  \hfill
  \begin{subfigure}[c]{0.84\textwidth}
      \centering
      \includegraphics[width=\textwidth]{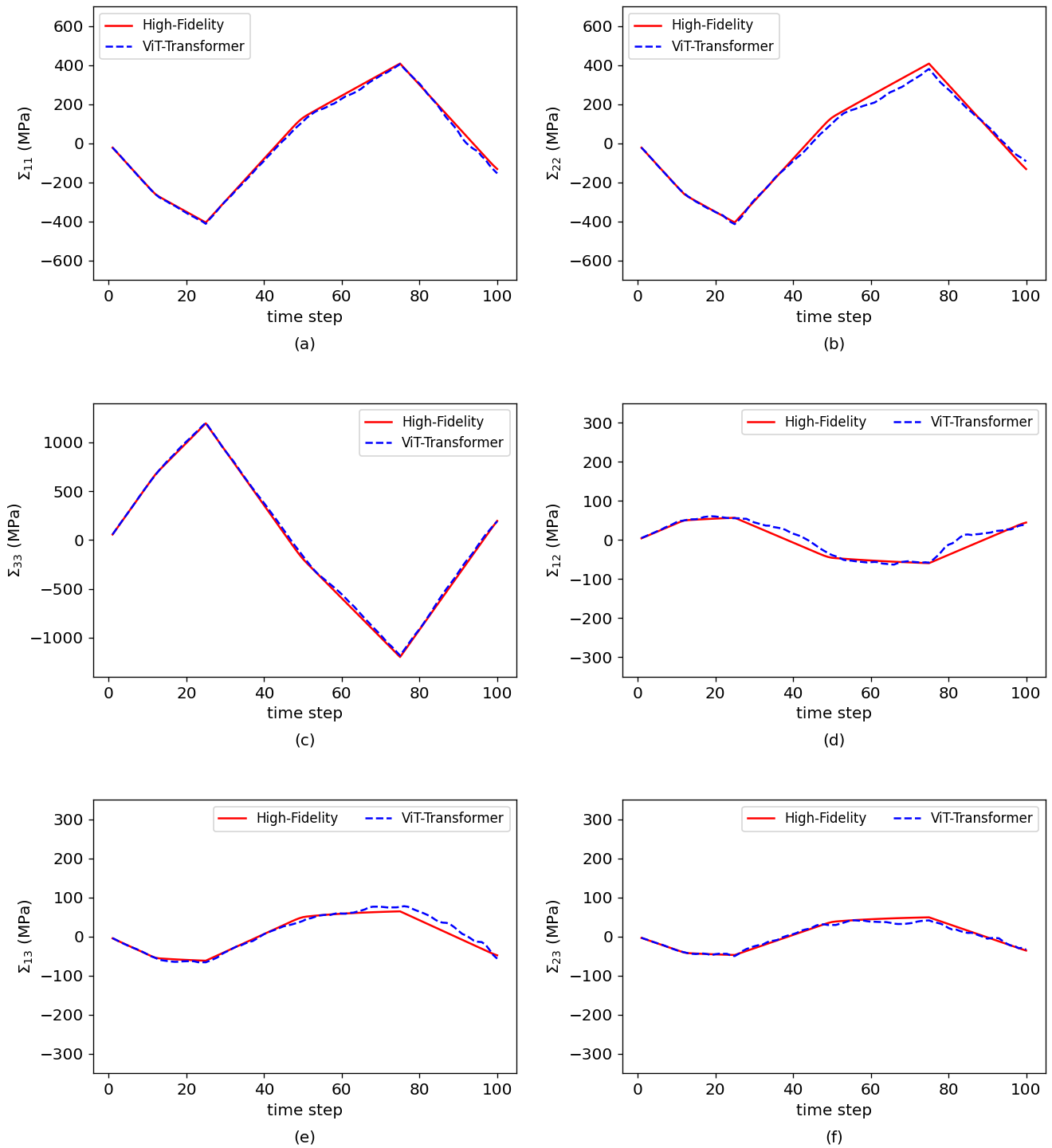}
      \label{fig:Test_in_unseen_3_2}
  \end{subfigure}
  \caption{Comparison of the predicted and true stresses corresponding to the unseen linear cyclic strain loading protocol with non-zero shear shown in Figure \ref{fig:triaxial_states}c. The 2D cross-sectional image of the microstructure is shown on the left. (a-f) Predicted and true macroscopic stresses versus time steps. Reference values are shown in red and blue dashed lines show the predicted values.}
  \label{fig:Test_in_unseen_3}
\end{figure}

In the case of linear cyclic protocol with non-zero shear strains (Figure \ref{fig:triaxial_states}c), $E_{11}$ and $E_{22}$ reach the peak value of $\pm$0.005 with $\pm$0.0002 increments. $E_{33}$ reaches the peak strain of $\pm$0.01 with fixed increment size of $\pm$0.0004. There are 100 time steps in this protocol.
The increments of shear strain components are $\Delta E_{12} = 0.0001$, $\Delta E_{13}=-0.0001$, and $\Delta E_{23} = 0.000075$. The geometrical features in this example are the same as Group 4 in Table \ref{tab:Dataset} with randomly positioned fibers (Figure \ref{fig:Test_in_unseen_3}). Figure \ref{fig:Test_in_unseen_3} shows the comparison of stress time history predicted by the ViT-Transformer model with the reference values,  leading to a relative error of 4.76\%.

\begin{figure}[h]
  \centering
  \begin{subfigure}[c]{0.15\textwidth}
      \centering
      \includegraphics[width=\textwidth]{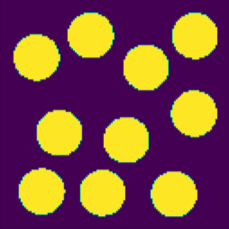}
      \label{fig:Test_in_unseen_4_1}
  \end{subfigure}
  \hfill
  \begin{subfigure}[c]{0.84\textwidth}
      \centering
      \includegraphics[width=\textwidth]{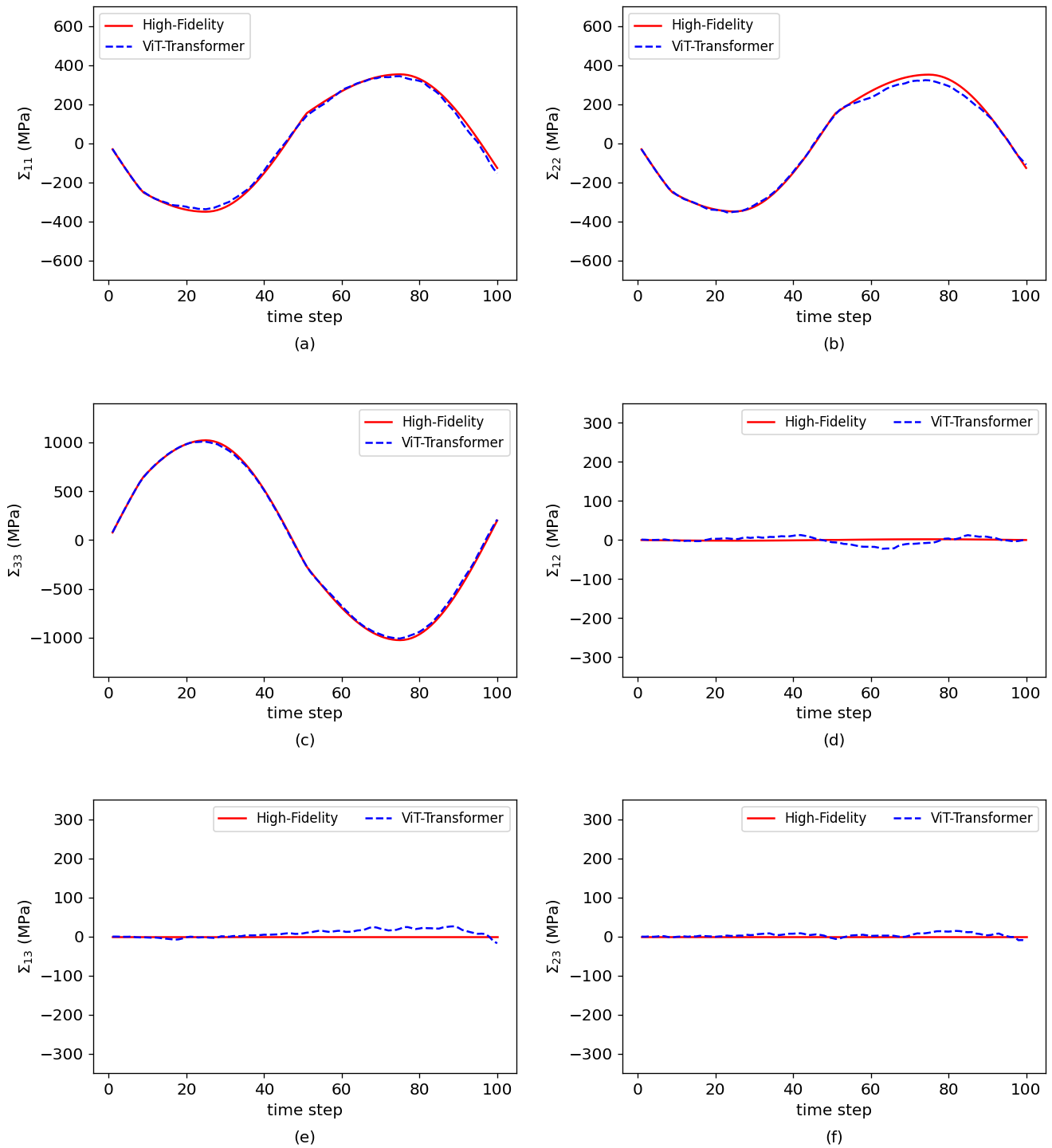}
      \label{fig:Test_in_unseen_4_2}
  \end{subfigure}
  \caption{Comparison of the predicted and true stresses corresponding to the sinusoidal loading protocol shown in Figure \ref{fig:triaxial_states}d. The 2D cross-sectional image of the microstructure is shown on the left. (a-f) Predicted and true macroscopic stresses versus time steps. Reference values are shown in red and blue dashed lines show the predicted values.}
  \label{fig:Test_in_unseen_4}
\end{figure}

\begin{figure}[h]
    \centering
    \includegraphics[width=12cm]{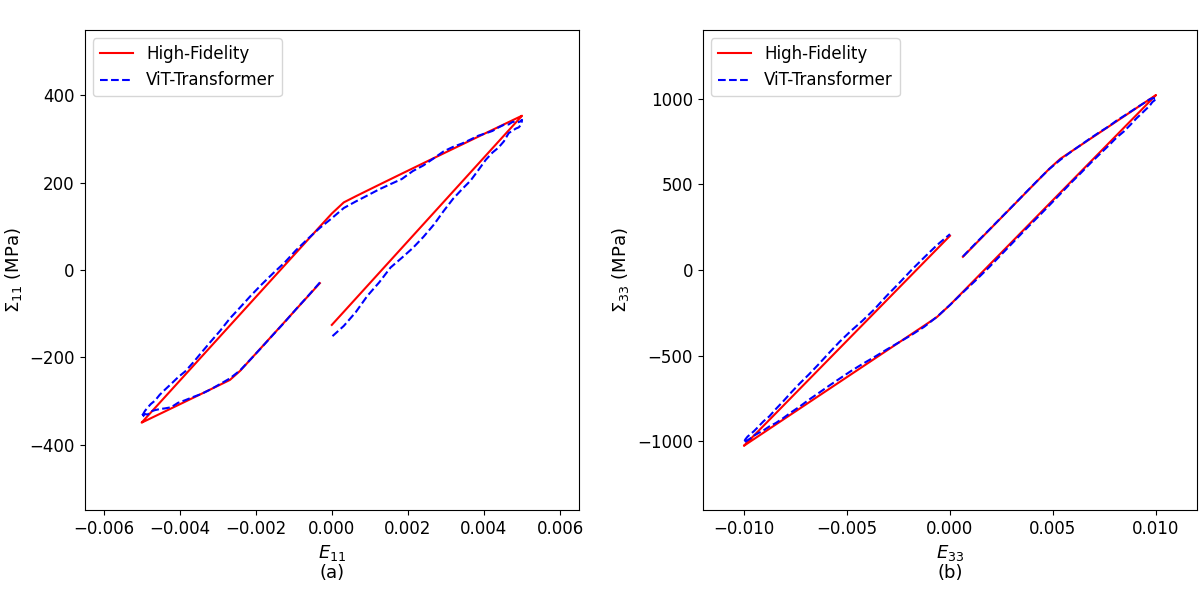}
    \caption{The reference and predicted macroscopic stress versus strain corresponding to the sinusoidal load shown in Figure \ref{fig:triaxial_states}d.}
    \label{fig:Test_in_unseen_4_3}
\end{figure}

In the sinusoidal loading case (Figure \ref{fig:triaxial_states}d), there are 100 time steps but the strain increments are not constant. $E_{11}$, $E_{22}$, and $E_{33}$ reach the peak values of $\pm$0.005, $\pm$0.005, and $\pm$0.01, respectively.
The geometrical features in this example are the same as Group 5 in Table \ref{tab:Dataset}, and the positions of fibers are randomly generated and unseen. Figures \ref{fig:Test_in_unseen_4} and \ref{fig:Test_in_unseen_4_3} show the model performance in predicting stresses with a relative error of 4.05\%. 

Overall, we observe from Figures  \ref{fig:Test_in_unseen_1} to \ref{fig:Test_in_unseen_4_3} that the macroscopic stresses predicted by the ViT-Transformer model follow the results of high-fidelity FEM simulations closely with small relative errors. We note that these loading protocols are different from the random-walk based training dataset; therefore, they are considered difficult for the model. In loading scenarios with zero shear stresses, the model correctly predicts zero values with small fluctuations. 

\subsubsection{Performance on  unseen microstructure}

\begin{figure}[h]
  \centering
  \begin{subfigure}[c]{0.15\textwidth}
      \centering
      \includegraphics[width=\textwidth]{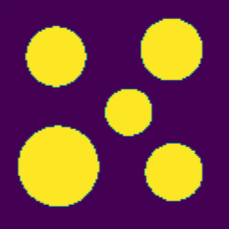}
      \label{fig:Test_in_unseen_5_1}
  \end{subfigure}
  \hfill
  \begin{subfigure}[c]{0.84\textwidth}
      \centering
      \includegraphics[width=\textwidth]{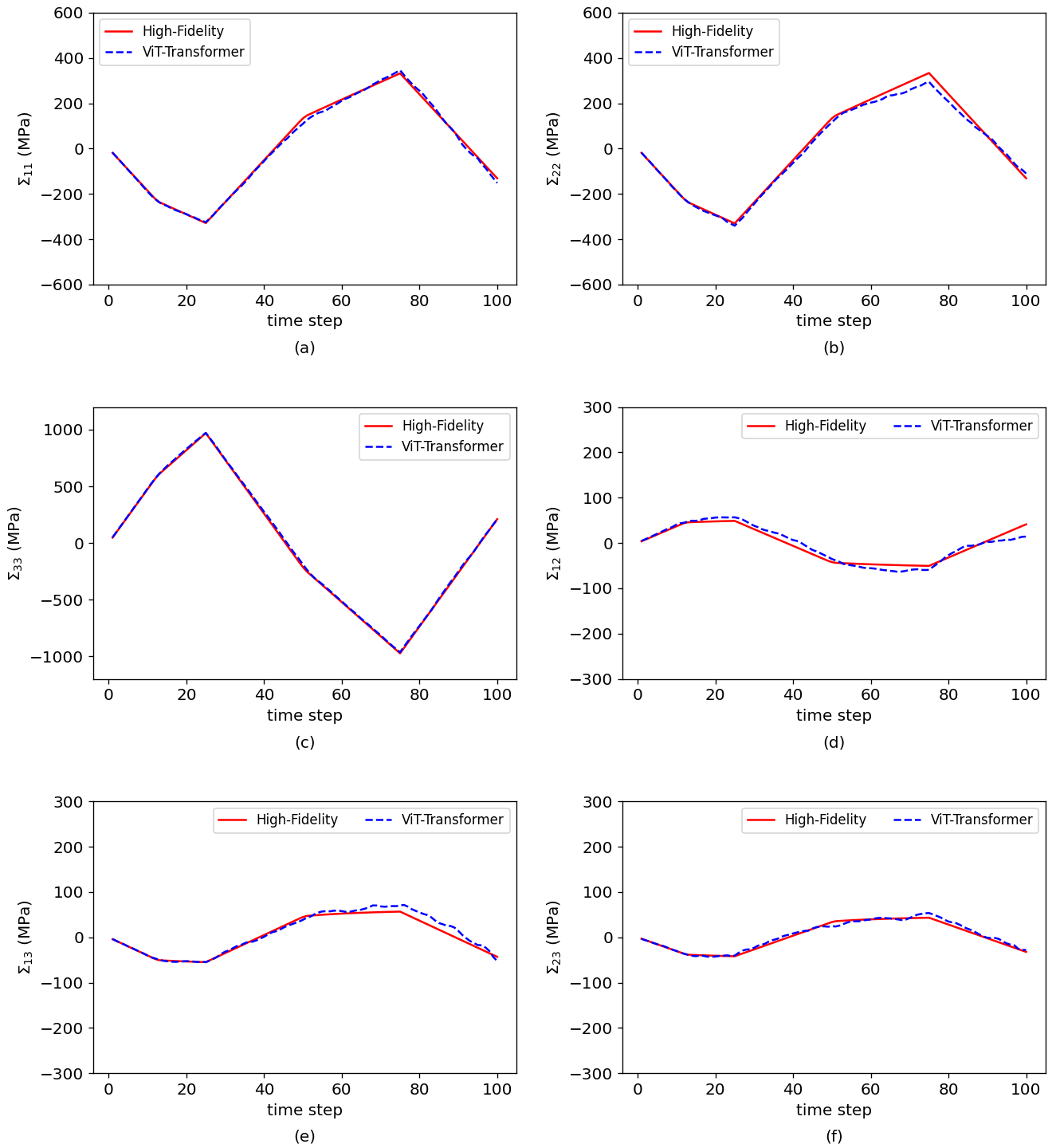}
      \label{fig:Test_in_unseen_5_2}
  \end{subfigure}
  \caption{Comparison of the predicted and true stresses corresponding to the unseen linear cyclic strain loading protocol with non-zero shear shown in Figure \ref{fig:triaxial_states}c. The 2D cross-sectional image of the unseen microstructure with randomly generated fiber radii and positions is shown on the left. (a-f) Predicted and true macroscopic stresses versus time steps. Reference values are shown in red and blue dashed lines show the predicted values..}
  \label{fig:Test_in_unseen_5}
\end{figure}

To evaluate the generalization capability of the ViT-Transformer model, we generate a more complex microstructure which has a completely different microstructural pattern from the training and testing datasets. In this composite RVE, the fiber positions and radii are both independently sampled randomly, as illustrated in Figure \ref{fig:Test_in_unseen_5}. The RVE is then subjected to the cyclic loading protocol shown in Fig. \ref{fig:triaxial_states}c. Using the same trained surrogate model as in the previous experiments, we predict the macroscopic stress sequence of this RVE by supplying the 2D cross-sectional image of the RVE together with the prescribed strain sequence. The performance of our model in predicting stresses compared to the high-fidelity simulation results is shown in Figure \ref{fig:Test_in_unseen_5}. The relative error in this case is 4.21 \%, which is similar to the numerical tests in Section \ref{sec:unseen_load}.

\subsection{Performance of the RET algorithm}\label{subsec:The Performance of Random Extract Training Algorithm}

\begin{figure}[h]
    \centering
    \includegraphics[width=0.84\textwidth]{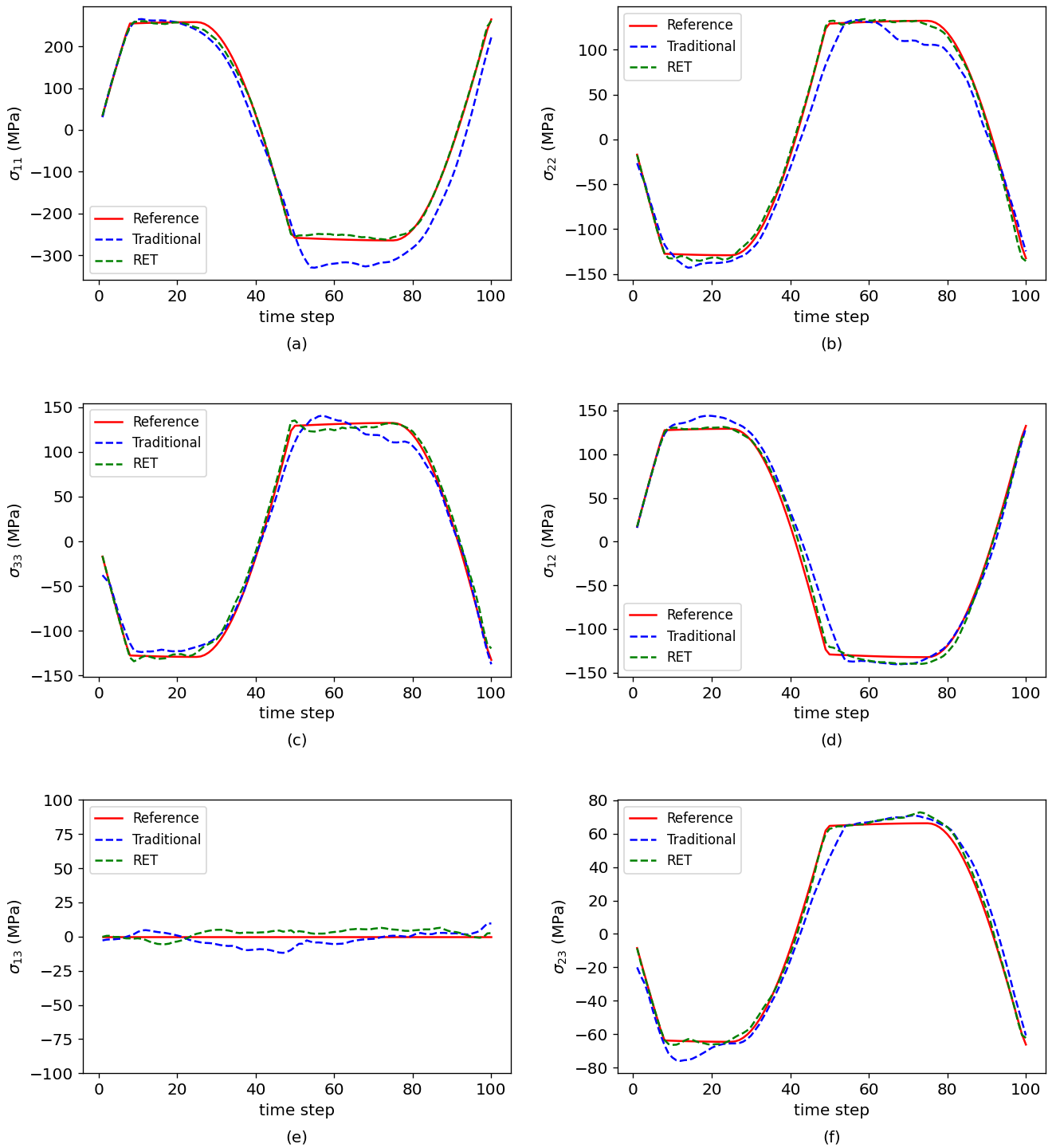}
    \caption{Comparison of the stress sequences predicted using Transformers trained by traditional (dashed blue line) and RET (dashed green line) algorithms. The reference solution is obtained using J2 elasto-plasticity (solid red line).}
    \label{fig:Training_comp}
\end{figure}

In this subsection, we investigate the effect of training algorithms on the surrogate model performance. Since the RET algorithm is only applicable to the decoder, this test targets the decoder alone. Accordingly, we generate a 3D dataset of stress and strain pairs using J2 elasto-plasticity without any microstructure. The dataset consists of 10,000 sequences, each of length 200. 90\% of the data is used for training and 10\% for testing. 
We train two Transformer decoders independently. One model is trained using the conventional approach, where the sequence length remains constant at 200 during training. The second model is trained using the RET algorithm. The hyperparameters associated with the model architectures are tuned and are selected to be the same for both models, as shown in Table \ref{tab:ViTT_HP}. The training-related hyperparameters (e.g., learning rate) were kept identical to those employed in the previous subsections. The training MSE losses of the two models reach around $8 \times 10^{-8}$ and test MSE losses for both models are around $10^{-7}$, showing that the models are well-trained. 
Two unseen strain sequences are generated with 100 and 200 time steps following the sinusoidal pattern shown in Figure \ref{fig:triaxial_states}d, with the values of ${E}_{12}$ and ${E}_{23}$ set equal to ${E}_{11}$ and ${E}_{22}$, respectively. Table \ref{tab:Training_comp} shows the relative error calculated using Equation \ref{eq:relative-error}. Figure \ref{fig:Training_comp} compares the predicted and reference stress time series for the sequence with length of 100.

In Figure \ref{fig:Training_comp}, the red line represents the true (reference) stress values, the blue dashed line indicates the predictions of the Transformer model trained with the traditional training algorithm, while the green dashed line represents the predictions of the Transformer model trained with the RET algorithm. We observe from Figure \ref{fig:Training_comp} that the RET algorithm significantly outperforms the traditional training algorithm. As indicated in Table \ref{tab:Training_comp}, when processing sequences of length 100, the model trained with the RET algorithm exhibits a relative error of one order of magnitude smaller compared to the model trained with the conventional training approach. Table \ref{tab:Training_comp} demonstrates that the performance of the model trained with the RET algorithm remains the same when processing sequences of different lengths. In contrast, the performance of the model trained with the conventional approach significantly deteriorates when processing sequences of length different from the one used in the training dataset.

\begin{table}[h]
\centering
\caption{Comparison of the relative errors of the Transformers trained by traditional and RET algorithm.}
\label{tab:Training_comp}
\renewcommand{\arraystretch}{1.5}
\begin{tabular}{l l l l}
\hline
Length of Sequences & Traditional & RET Algorithm \\[5pt]\hline
200 & 0.063 & 0.047 \\[5pt]
100 & 0.168 & 0.045 \\[5pt]\hline
\end{tabular}
\end{table}

\subsection{Comparison of Transformer-based and GRU surrogate models}

In this section, we compare the capability of Transformer-based surrogate model in capturing long-range dependencies with its GRU-based counterpart. We focus on the performance of the self-attention mechanism, therefore, we train both models on the same dataset employed in Section \ref{subsec:The Performance of Random Extract Training Algorithm}. Since this test has no microstructure to process, we only use the decoder part of ViT-Transformer.
We have performed a hyperparameter search for GRU and Transformer based decoder with the new dataset, and we have found that the hyperparameters adopted in our earlier work \cite{zhou2025machine} for GRU, as shown in Table \ref{tab:GRU_HP}, and those reported in Section \ref{sec:methodology} for the Transformer decoder are optimal. Only the hyperparameters associated with the model architecture were tuned, whereas training-related hyperparameters (e.g., learning rate) were kept identical to those employed in the previous sections. The ultimate training and testing MSE losses of the Transformer model are about $7.5 \time 10^{-8}$ and $10^{-7}$, respectively. For the GRU model, the training and testing MSE losses are around $10^{-7}$ and $1.5 \times 10^{-7}$, respectively, showing that both models are trained well.

\begin{table}[h]
\centering
\caption{Hyperparameters used in the GRU model}
\begin{tabular}{l l l}
    \hline 
    Unit & Parameters & Value \\[2pt]
    \hline
     & Number of layers & 3 \\
     GRU & Number of inputs & 6 \\
     & Number of hidden states & 50 \\[5pt]
      & Number of inputs & 50 \\
   FC layer  & Number of outputs & 6 \\
     & Number of layers & 1 \\[5pt]
    \hline
\end{tabular}
\label{tab:GRU_HP}
\end{table}

\begin{figure}[h]
    \centering
    \includegraphics[width=15cm]{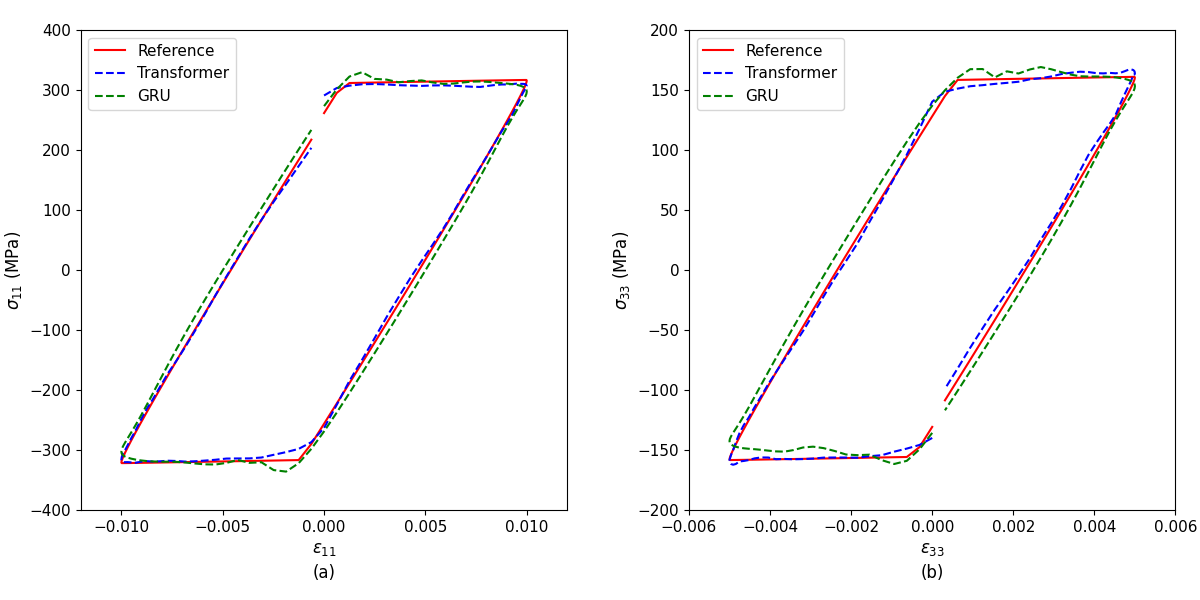}
    \caption{Comparison of the predicted stress versus strain response obtained from the Transformer (blue dashed line) and GRU (green dashed line) models under triaxial sinusoidal cyclic strain with zero shear strain (steps 101 to 200). The red line represents the reference value. }
    \label{fig:Compa_T_G}
\end{figure}

We extend the same strain sequence from Figure \ref{fig:triaxial_states}d to 200 steps to test the Transformer and GRU surrogate models. Figure \ref{fig:Compa_T_G}, compares the predictions of the Transformer and GRU models from step 101 to step 200 for $\sigma_{11}$ and $\sigma_{33}$ versus the corresponding strain components. We observe that the predictions of the Transformer surrogate model are more accurate and stable compared to the GRU model. Table \ref{tab:Compa_T_G} compares the relative errors of the Transformer and GRU models for the the first and second halves of the loading sequence, showing that the Transformer model provides more accurate results compared to GRU for both halves of the sequence.

\begin{table}[h]
\centering
\caption{Comparison of the relative errors of the Transformer and GRU models over the entire prediction process and in the latter part of the predictions.}
\label{tab:Compa_T_G}
\begin{tabular}{l c c}
\hline
 & GRU & Transformer \\[5pt]\hline
Steps 1 to 100 & 0.054 & 0.036 \\[5pt]
Steps 101 to 200 & 0.056 & 0.034 \\[5pt]\hline
\end{tabular}
\end{table}

\section{Concluding remarks}\label{sec:conclusion}

In this work, we presented ViT-Transformer, a novel surrogate constitutive model based on the self-attention mechanism for predicting the path-dependent macroscopic response of heterogeneous materials. The model combines a Vision Transformer encoder developed for capturing microstructural information from RVE images with a masked Transformer decoder to process the input strain sequence and predict the macroscopic stress response.
We demonstrated the application of this model for 
three-dimensional long-fiber composites. 
To reduce computational expenses resulting from microstructure embedding, we utilized two-dimensional images of composite RVEs taken in planes perpendicular to the fiber direction as model inputs to capture microstructural features, which were then leveraged for three-dimensional surrogate constitutive modeling of composites.
In addition, we developed a random extract training algorithm that exposes the model to sequences of different lengths during training to further enhance its capability to generalize to sequences with variable lengths. We designed a compact yet information-rich dataset with the aid of data augmentation.
We carefully evaluated the model performance on unseen microstructures and under monotonic, cyclic, and sinusoidal loading which were all unseen during the training process. Comparison with high-fidelity FEM simulations shows that the proposed model captures the history-dependent behavior of the composite RVEs accurately and generalizes well across different microstructures. 

This work demonstrates the robustness of self-attention mechanism in capturing the nonlinear elasto-plastic response of composite materials with varying microstructure. 
We showed that ViT-Transformer architecture outperforms the standard GRU based architectures in capturing long-range dependencies in longer loading sequences. Moreover, by employing attention mechanisms for both long-sequence modeling and image-based microstructural feature processing, ViT-Transformer can generalize well to unseen microstructures and path-dependent loading protocols. The results suggest that attention has a promising potential for future research in multiscale simulations of composites and heterogeneous materials.

\end{document}